\documentclass[preprint]{elsarticle}

\usepackage[utf8]{inputenc}
\usepackage[T1]{fontenc}
\usepackage{hyperref}
\usepackage{lscape}
\usepackage{graphicx}
\graphicspath{{IMAGES/}}
\usepackage{gensymb}
\usepackage{geometry}
\usepackage{longtable}
\usepackage{amsmath}
\usepackage{amssymb}
\usepackage{xcolor}
\usepackage{eso-pic}
\usepackage{url}

\AddToShipoutPictureFG{%
  \AtPageUpperLeft{%
    \put(\LenToUnit{\dimexpr\paperwidth-7mm\relax},%
         \LenToUnit{-0.62\paperheight}){%
      \rotatebox{90}{\sffamily\footnotesize\color{gray}%
      ACCEPTED by Earth and Planetary Science Letters, July 2026}%
    }%
  }%
}

\newcommand{\ins}[1]{#1}
\newcommand{\del}[1]{}
\newcommand{\linkref}[2][]{#1~#2}
\newcommand{\hlsection}[1]{}
\newcommand{\PrintCredit}{}
\newcommand{\linkto}[1]{}
\newcommand{\elsalttext}[1]{}
\providecommand{\xsnm}[1][]{#1}
\providecommand{\xfnm}[1][]{#1}
\providecommand{\PrintOrdinal}[1]{#1}
\newcommand{\appsection}[1]{\section{#1}}
\newenvironment{appgroup}{\appendix}{}

\newenvironment{ack}{}{}

\def\degree{^{\circ}}

\journal{Earth and Planetary Science Letters}
\begin{document}

\begin{frontmatter}

\title{The dynamical surface of Phobos: A morphodynamic atlas}

\author[add1,add3]{Isabel Herreros\corref{mycorrespondingauthor}}
\ead{iherreros@cab.inta-csic.es}

\author[add2]{S\'ebastien Charnoz\corref{mycorrespondingauthor}}
\ead{charnoz@ipgp.fr}

\cortext[mycorrespondingauthor]{Corresponding authors.}

\address[add1]{Centro de Astrobiolog\'{\i}a CSIC-INTA, Carretera Ajalvir Kil\'ometro 4, 28850 Torrej\'on de Ardoz, Madrid, Spain}

\address[add2]{Universit\'e Paris Cit\'e, Institut de Physique du Globe de Paris, CNRS, 75005, Paris, France}

\address[add3]{Departamento de Ingenier\'{\i}a T\'ermica y Fluidos, Universidad Carlos III de Madrid, 28911 Legan\'es, Madrid, Spain}

\begin{abstract}
Phobos evolves in a highly dynamical environment where surface-material motion is controlled by the combined effects of self-gravity, time-dependent Martian tides, and inertial forces. In such a low-gravity regime, the displacement of loose material, like regolith, cannot be inferred from topographic slope alone, making a dynamical approach essential for interpreting Phobos' surface morphology and for supporting the Martian Moons eXploration (MMX) mission led by JAXA. Here, using our RAVEL code, we apply a dynamical model that combines the surface acceleration field, including gravity, centrifugal and tidal components, with friction on a digital terrain model of Phobos to compute surface regolith trajectories. The model does not aim to predict the triggering of slope failure. Instead, it addresses where material would preferentially move once motion is initiated. This reveals large-scale coherent dynamical regions and a sparse network of preferred regolith transport routes, termed here \textit{Regolith Migration Pathways} (RMPs).
The final positions of the RMPs correlate with smooth, low-relief terrains and spectrally neutral units, consistent with depositional mantles formed by long-term regolith infill, whereas rough, high-standing areas with abundant small craters and blue spectral slopes tend to correspond to dynamically active or denuded source regions. In contrast, spectrally red terrains are generally associated with dynamically quiet, morphologically rough surfaces where our model predicts negligible regolith motion, suggesting older, less frequently reworked units. Taken together, these patterns indicate that much of Phobos' surface morphology and spectral heterogeneity can be explained by long-term regolith redistribution driven by the surface acceleration field along RMPs. We provide a 3D morphodynamic atlas of RMPs across Phobos' surface, which will be useful for constraining the geographical provenance of samples to be collected by the MMX spacecraft.
\end{abstract}


\end{frontmatter}

\section{Introduction}\label{Xsec1-1}

Phobos displays a wide range of surface features, including grooves, lineaments, landslides, { and relatively smooth, sparsely cratered terrains at image scale,} as well as different spectral units, whose origins are still debated. Proposed formation mechanisms, such as tectonic fracturing, ejecta emplacement from Stickney crater, seismic shaking due to impacts or tides or secondary impacts, have been proposed but not conclusively validated  (see e.g. \citealp{Veverka78, Wahlisch_2010, Basilevsky_2014, Ballouz2019, ramsley2021}). Evidence for mass wasting, including avalanches and downslope regolith motion on crater walls, further suggests that surface material can be mobilized under present or past conditions \citep{Basilevsky_2014, shi2013}.
In Phobos' low-gravity environment, stability and mobility of surface material are controlled not only by local topography but also by orbital and rotational forces (centrifugal, tidal, and Coriolis). Several studies have mapped the resulting \textit{dynamical slope} (\linkref[\del{equation}\ins{Eq.}]{\ref{dyn_slope}}) across the surface \citep{Ballouz2019, Ernst_2023}, providing useful insights into potential instability zones. However, such static analyses cannot capture the true paths followed by the mobile regolith, as trajectories are generally not simply aligned with local slopes and are also influenced by spatially and temporally varying tidal forces.

To understand how regolith is redistributed across Phobos' surface, it is therefore necessary to model the dynamical motion of surface material under the combined action of gravity, tides, rotation, and also friction. In this work, we address this gap by (1) analyzing in detail the surface force field at Phobos and (2) modeling regolith motion at the global scale, allowing us to identify coherent dynamical regions and preferential pathways of surface material transport.

The aim of this paper is not to address or determine the processes that trigger the initial failure of a slope and the onset of regolith motion. Such triggering may depend on local and poorly constrained processes, such as meteoritic impacts, fracture opening, seismic shaking, thermal stresses, or electrostatic effects. Instead, we ask a different question: if surface material becomes mobile, where does it move, along which transport routes, and where does it accumulate? A region can be dynamically quiet at a given time and still be a preferential transport corridor if the surface is later disturbed. We refer to these preferred transport routes as \textit{Regolith Migration Pathways}. To our knowledge, this is the first global calculation of such pathways over the surface of\break Phobos.

Therefore, in this work we focus on the routing of regolith at the surface and then investigate possible correlations with observed surface features, such as the red-blue spectral units and crater-poor versus crater-rich regions. To do so, we compute surface-constrained trajectories, i.e., trajectories that do not detach from the surface, using a Coulomb-type friction law, in which the effective friction angle is treated as a \textit{mobility-threshold criterion}. This strategy is motivated by the fact that both the paleotopography at the time of past transport episodes and the mechanical properties of the regolith remain unconstrained: rather than prescribing speculative ancient topographies and intrinsic friction properties, we explore two end-member values of the effective friction angle within the present-day topography. The lower end-member, $14^\circ$, is close to the present-day mean dynamical slope of Phobos and is adopted to reveal the broader underlying transport corridors, which may be representative either of past episodes of spontaneous regolith redistribution or of present-day transport initiated by external perturbations such as seismic shaking, nearby impacts, fracture-related disturbances, or thermal stresses. The higher end-member, $30^\circ$, provides a more conservative reference case for identifying regions that may still be compatible with spontaneous regolith motion under present-day conditions. This approach allows us to identify dynamically active regions and preferential \textit{Regolith Migration Pathways} across the surface. We use these pathways to interpret surface morphologies and spectroscopic units, assess terrain stability and resurfacing potential, and link local observations to the large-scale evolution of Phobos' surface.

This analysis is particularly timely in the context of the forthcoming Martian Moons eXploration (MMX/JAXA) mission, which will return surface samples from Phobos and requires a well-constrained dynamical basis to interpret their provenance. By identifying regions where regolith transport and surface instability may be favored, this work directly supports future mission planning and contributes to the evaluation of candidate landing and sampling sites. Furthermore, by linking surface trajectories to possible source regions, it helps constrain the provenance of collected material and strengthens the geological interpretation of the returned samples.

The paper is organized as follows. \linkref[Section]{\ref{section2}} describes the computation of the surface acceleration field on Phobos and introduces a three-dimensional representation of surface acceleration vectors. \linkref[Section]{\ref{section:accel_dynamical_regions}} presents the resulting surface dynamical features. In \linkref[Section]{\ref{Section4}}, we compute surface-constrained trajectories  and identify a network of \textit{Regolith Migration Pathways} (RMPs). \linkref[Section]{\ref{results}} compares these pathways with spacecraft observations, including Mars Express/HRSC and MRO imagery and spectral slope maps. \linkref[Section]{\ref{Conclusion}} discusses the implications of these results for surface evolution on Phobos.

\section{Acceleration at the surface of Phobos}\label{Xsec2-2} \label{section2}

The computation of surface accelerations on Phobos is described in \linkref[Appendix]{\ref{Appendix: acceleration}} and follows standard approaches \citep{Ballouz2019, Ernst_2023}. Phobos is assumed to be on its present eccentric orbit \citep{Jacobson_Lainey_2014}. {It is assumed to have a homogeneous interior, and to be in synchronous rotation with constant rotation velocity  (equal to its mean-motion). Phobos libration, about} 1.14$^{\circ}$ \citep{Rambaux_2012}, is not included. Unless otherwise stated, Phobos is placed at pericenter (we checked that the eccentricity has a minor effect on our results, see discussion in \linkref[Appendix]{\ref{subsec:slope_stability}}). { In the computation of the acceleration field, all acceleration terms are included except for the Coriolis term, which depends on the instantaneous velocity and is therefore introduced only after the velocity has been computed} (see \linkref[Appendix]{\ref{Appendix:trajectories}}). Calculations are performed using a 36\del{ }\ins{\thinspace }m resolution digital terrain model of Phobos \citep{Ernst_2023}.

Previous studies have typically characterized local surface forcing using the local dynamical slope, defined as the angle between the surface normal and the acceleration vector. While useful, this scalar quantity discards most directional information. Here, we instead project the local acceleration onto the plane tangential to the local topography, which provides a clearer representation of acceleration directions and enables the identification of large-scale dynamical regions across Phobos' surface. They are presented in the next section.

\section{Identification of dynamical features on Phobos from the surface acceleration field}\label{Xsec3-3}
\label{section:accel_dynamical_regions}

\begin{figure*}

\centerline{\includegraphics[width=0.95\textwidth]{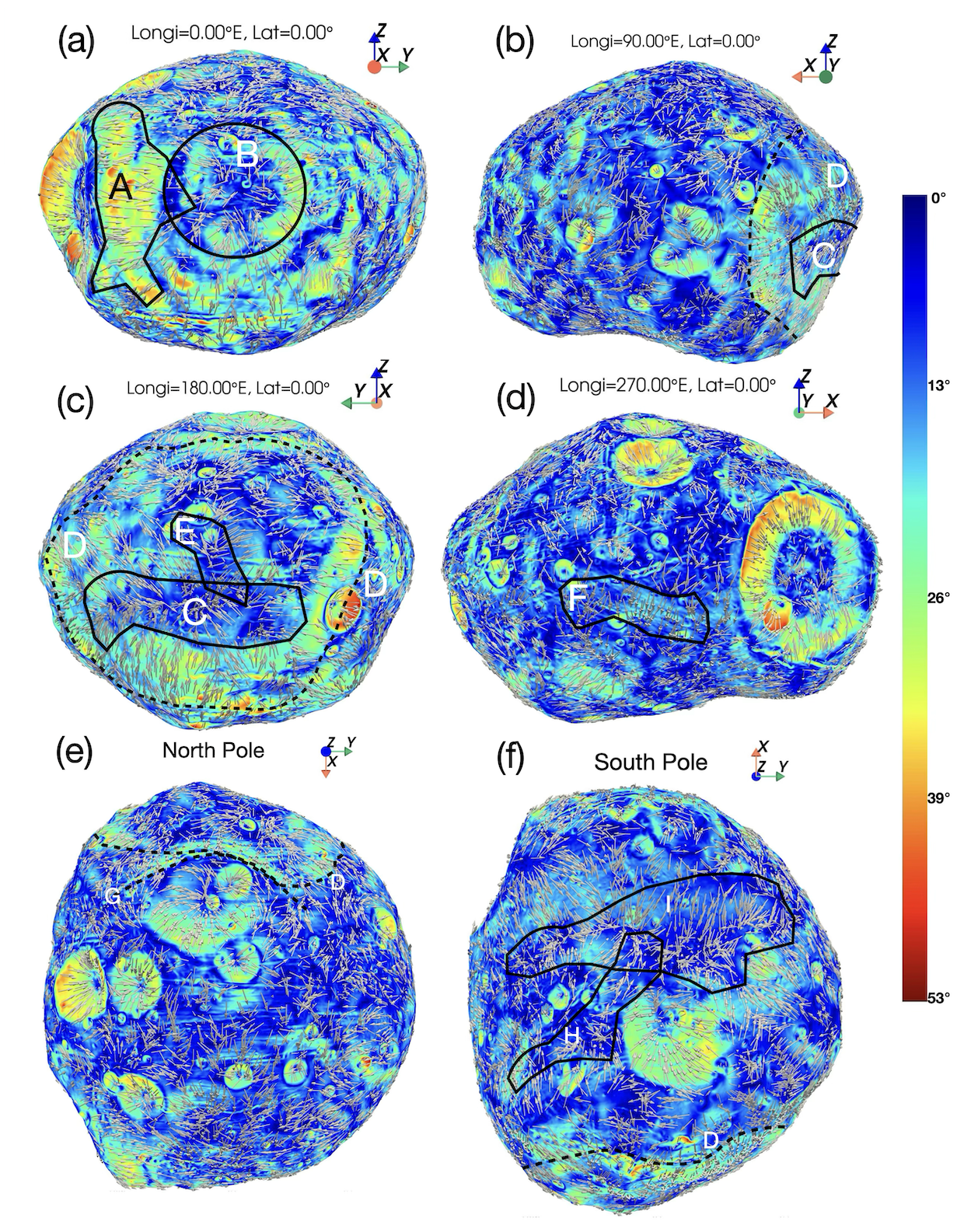}}
\caption{Surface acceleration map of Phobos. The colors represent the local dynamical slope (color bar on the right). The arrows show the direction of the local acceleration field projected onto the topography (\( \vec{a}_{\parallel} \)), \linkref[\del{Equation}\ins{Eq.}]{\eqref{equation:a_surface}}. The length of the arrows is proportional to the magnitude of the tangential acceleration. Phobos is rotated around the Z axis. The X direction points toward Mars, the Y direction is opposite to Phobos' motion, and the Z direction aligns with Phobos' rotation axis.   Black lines delineate regions (features A to I) such as depressions or troughs (solid line) or rims and ridges (dashed line). See \linkref[\del{section}\ins{Section}]{\ref{section:accel_dynamical_regions}} for details.}
    \label{figure:accel_map_6panels_FEATURES}
\end{figure*}

We plot in \linkref[\del{Figures}\ins{Fig.}]{\ref{figure:accel_map_6panels_FEATURES}}a to d the acceleration maps along the equator, centered at east longitudes 0$^\circ$ (sub-Mars point), +90$^\circ$, +180$^\circ$ (anti-Mars point) and +270 $^\circ$. \linkref[\del{Figure}\ins{Fig.}]{\ref{figure:accel_map_6panels_FEATURES}}e and f present the north and south poles. Regions showing a specific organization of acceleration vectors (in orientation and magnitude) define dynamical features on the surface of Phobos. Some of these features can be clearly associated with topographic structures (e.g., along the edge of Stickney crater), whereas others are revealed only through a detailed analysis of the acceleration maps (e.g., near the sub-Mars point). The most prominent features are delineated by black lines in \linkref[\del{Figure}\ins{Fig.}]{\ref{figure:accel_map_6panels_FEATURES}}.

\begin{itemize}
    \item \textbf{Dynamical Feature A: } \textit{Eastern Outer Slope of Stickney Crater } (\linkref[\del{Figure}\ins{Fig.}]{\ref{figure:accel_map_6panels_FEATURES}}a): This is a well-known region of Phobos, where the external slope of the eastern rim of Stickney crater corresponds to one of the steepest areas on this moon. The acceleration vectors are directed eastward and point toward topographic Feature~B (Sub-Mars \textit{dynamical depression}). The acceleration field aligns qualitatively with the grooves observed along the eastern outer slope of Stickney crater, suggesting that these grooves were likely carved by material sliding downslope from the Stickney region, as suggested in \cite{Ramsley_2019}. See detailed discussion in \linkref[\del{section}\ins{Section}]{\ref{section_east_of_stickney_VS_images}}.

    \item \textbf{Dynamical Feature B:} \textit{Sub-Mars Low }(\linkref[\del{Figure}\ins{Fig.}]{\ref{figure:accel_map_6panels_FEATURES}}a). Feature B is circular in shape. { Acceleration vectors  along the outer edges of the feature (where slopes are steeper) consistently point toward the center, indicating a convergent pattern. In contrast, the central region corresponds to a relatively flat area (low slope, shown in dark blue), where the gradient is weak and vector directions become less coherent and  more scattered.}

    \item \textbf{Dynamical Features C and E:} \textit{Anti-Mars Southern Troughs} (\linkref[\del{Figure}\ins{Fig.}]{\ref{figure:accel_map_6panels_FEATURES}}b and c). Feature C corresponds to a remarkable, narrow  and elongated region located in the southern hemisphere, southwest of the anti-Mars point. It is a low-slope area where acceleration vectors converge and align with the local trough orientation. It is connected to a narrower and shorter trough-like structure, Feature~E, which encompasses the anti-Mars point itself. According to our acceleration map, this area corresponds to a convergence zone with linear structure. See our detailed comparison with Mars Express data in \linkref[\del{section}\ins{Section}]{\ref{section_discussion_southern_through_comparison_with_images}}.

    \item \textbf{Dynamical Feature D:} \textit{Far-side Annular Ridge} (\linkref[\del{Figure}\ins{Fig.}]{\ref{figure:accel_map_6panels_FEATURES}}c). This feature appears as an arcuate, nearly circular dynamical structure extending across a large portion of Phobos' far side. It is defined by a broad zone where acceleration vectors change orientation, forming an apparent annular pattern. Although its morphology may resemble a circular ridge, no clear topographic counterpart has been identified in\break existing shape models, and it is therefore interpreted here as a large-scale dynamical boundary.

     \item \textbf{Dynamical Feature F:}\textit{ Leading Side Southern Low  }(\linkref[\del{Figure}\ins{Fig.}]{\ref{figure:accel_map_6panels_FEATURES}}d) Feature F corresponds to an elongated region near longitude 270° E, west of Stickney, where both the dynamical slope and the topography indicate a local low. The acceleration vectors within this area exhibit a slightly convergent pattern toward the center of the feature, consistent with a minimum in the effective acceleration field. Such a configuration may promote regolith retention or limited downslope motion within this western depression. This feature seems to connect to Feature I (south-pole trough).

    \item \textbf{Dynamical Feature G:} \textit{ Grildrig Rim  }(\linkref[\del{Figure}\ins{Fig.}]{\ref{figure:accel_map_6panels_FEATURES}}e) Feature G is an arc-like structure tangent to the Grildrig Crater on the north pole.

     \item \textbf{Dynamical Features H and I:} \textit{South Pole Troughs} (\linkref[\del{Figure}\ins{Fig.}]{\ref{figure:accel_map_6panels_FEATURES}}f) Feature I is a remarkable, nearly linear depression located near Phobos' south pole, close to the anti-Mars hemisphere, with a length comparable to Phobos' diameter. The acceleration field is consistently aligned along the axis of this depression, defining a preferred downslope direction, while the central part of the structure exhibits very low slopes. Feature H is an elongated narrow trough that appears to connect with Feature I. Comparison with Mars Express data is discussed in\break \linkref[\del{section}\ins{Section}]{\ref{section_discussion_south_pole_comparison_with_images}}.

\end{itemize}

Of course, these acceleration maps provide information only about the direction of the local acceleration with respect to the local topography. This offers only a crude view of the actual regolith motion that may occur on Phobos' surface under the action of gravitational and centrifugal forces. To better constrain the occurrence and location of this motion, we now turn to a stability analysis and to the calculation of surface trajectories.

\section{Computation of surface trajectories}\label{Xsec4-4} \label{Section4}

Here we describe how surface trajectories are computed with \textit{RAVEL (Regolith Astrodynamics in Variable Effective Low-Gravity Environments)}, a numerical code developed by the authors to integrate the motion of surface tracers over irregular small-body topography under gravity, inertial forces, and friction. Full methodological details are provided in \linkref[Appendix]{\ref{Appendix:trajectories}}. { We adopt a kinematic approach in which we integrate the motion of fictitious massless tracers in the acceleration field that was presented in \linkref[\del{section}\ins{Section}]{\ref{section2}}}, in order to obtain a set of independent trajectories, referred to here as \textit{Regolith Migration Pathways}. The model includes self-gravity, Martian tidal forces, as well as centrifugal and Coriolis accelerations. Surface friction is implemented as a \textit{mobility-threshold criterion} that governs the initiation and cessation of motion, allowing us to identify Regolith Migration Pathways and their source and sink regions. We further assume that trajectories remain in continuous contact with the surface and do not allow for detachment, which could in principle occur during rapid motion over highly irregular terrain with strong local slope variations. Motion is constrained to the surface by removing the component of the velocity vector normal to the local topography.

Friction is represented using a Coulomb-type law. During motion, the friction force is
$
\mathbf{F}_{\mathrm{fric,dyn}} = -k_{\mathrm{dyn}}\,R\,\mathbf{u}$
where $k_{\mathrm{dyn}}$ is the dynamic friction coefficient, $\mathbf{u}$ is the unit vector in the direction of motion, and $R$ is the norm of the normal reaction force, \textbf{R}, exerted by the surface. The corresponding dynamic friction-angle is $k_{\mathrm{dyn}} =tan(\phi_{\mathrm{dyn}})$. In the static regime, the force required to maintain equilibrium must  remain smaller than the maximum static friction:
$
\|\mathbf{F}_{\mathrm{fric, static}}\| \le \tan(\phi_{\mathrm{static}})\,R,
$
where $\phi_{\mathrm{static}}$ is the static friction angle.

The friction coefficients used here should not be interpreted as { the intrinsic inter-grain friction of the regolith}. Rather, they define a { \textit{mobility-threshold criterion} used to identify dynamically active regions and preferential \textit{Regolith Migration Pathways} across the surface}. { The end-member values adopted in this work} are discussed in \linkref[Section]{\ref{section_surface_trajectories_and_different_friction_coefs}} and in the conclusion.

{
The normal reaction force $\mathbf{R}$, which enters the friction term, is computed under the assumption that, as long as the trajectory remains in contact with the surface, it balances the component of the applied acceleration field normal to the local topography. Hence, the magnitude of $\mathbf{R}$ scales with the component of the local effective gravity field normal to the surface. Since both the normal reaction force and the associated friction term depend on the same normal loading, the trajectory equations are formulated entirely in terms of accelerations. The proposed kinematic framework is therefore independent of any explicit mass scale.

More generally, the present approach is intentionally restricted to a purely kinematic framework. It does not attempt to resolve grain-scale mechanics or the detailed contact physics of the surficial material, such as inter-grain collisions, cohesion, rolling, or frictional interactions at the scale of individual grains or blocks. Instead, the trajectory calculations are formulated only in terms of the local acceleration field, the surface constraint, and a mobility-threshold criterion governing the initiation and cessation of motion. In that sense, the purpose of the model is not to reproduce the detailed physical behaviour of regolith, but to reveal the large-scale organization of \textit{Regolith Migration Pathways} across the surface of Phobos.

Within this framework, we focus on surface-constrained trajectories, which are computationally efficient and capture the dominant transport patterns. The validity of this approximation is discussed in}
\linkref[Appendix]{\ref{appendix:flying_particles_dynamics}},
{
where we compare these results with simulations that allow temporary detachment from the surface.
}

\section{Results}\label{Xsec5-5}
\label{results}

We present here the computed surface trajectories of regolith. We compare the predicted Regolith Migration Pathways, and their termination zones, with available imaging and spectral datasets to assess whether regions of  regolith motion exhibit distinctive textural or spectral signatures.

\subsection{Surface trajectories}\label{Xsec6-5.1}
\label{section_surface_trajectories_and_different_friction_coefs}
\begin{figure}

\centerline{\includegraphics[width=0.65\linewidth]{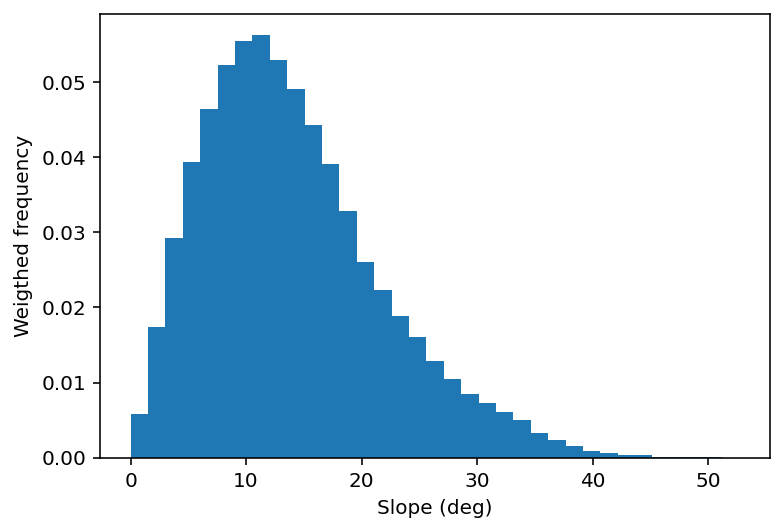}}
\caption{Histogram of slope distribution at the surface of Phobos. The mean slope is 14$^{\circ}$.}     \label{fig:slope_histrogram}
 \end{figure}

The  surface-constrained trajectories depend primarily on their initial location and on the static and dynamic friction angles,
$\phi_{\text{static}}$ and $\phi_{\text{dynamic}}$, both of which remain poorly constrained for Phobos.
In the absence of an initial velocity, downslope motion can occur only where the local dynamical slope exceeds the static friction threshold.
Considering that the mean surface slope of Phobos is only $14^{\circ}$ \citep{Ernst_2023} (\linkref[\del{Figure}\ins{Fig.}]{\ref{fig:slope_histrogram}}), adopting effective friction angles higher than $14^{\circ}$ would imply very limited surface mobility,
except within steep crater walls and scarps. Only these regions exhibit dynamical slopes above $30^{\circ}$, and they account for only a small fraction of Phobos' total surface area, as shown by the slope distribution in \linkref[\del{Figure}\ins{Fig.}]{\ref{fig:slope_histrogram}}.

It is, however, plausible that the present-day dynamical slopes of Phobos are significantly lower than in the past,
as progressive accumulation of a thick regolith layer has likely smoothed the topography, consistent with thermal inertia measurements \citep{Smith2018} { and with the very high surface porosity of Phobos' regolith, estimated to exceed 86\% \del{Fornasier {\rm et al.}, 2024)}} \citep{fornasier2024}.

{ Given that  both the paleotopography at the time of past transport episodes and the mechanical properties of the regolith remain unconstrained, we explored a suite of simulations for effective friction angles with the present-day topography ($14^{\circ}$, $20^{\circ}$, $25^{\circ}$, $30^{\circ}$, and $35^{\circ}$) finding that, although the overall geometry of surface trajectories remained consistent across cases,
their lengths systematically decrease with increasing friction. This indicates that the main transport corridors are largely controlled by the acceleration field and topography, whereas the friction angle primarily modulates the extent of transport along them.}

{ In this work we do not aim to reproduce the detailed behaviour of regolith on Phobos, but rather to identify the underlying transport corridors, i.e. the RMPs, that are compatible with the present-day acceleration field and topography. To this end, instead of prescribing speculative ancient topographies and frictional properties, we treat the effective friction angle as a \textit{mobility-threshold criterion} and consider two end-member values:}

(1) { The higher end-member, $30^\circ$, consistent with recent estimates  based on the morphology of rocks at the surface of Ryugu, Dimorphos and Bennu} \citep{robin2024}, { provides a more conservative reference case for identifying regions that may still be compatible with spontaneous regolith motion under present-day conditions;}
and (2) { the lower end-member, $14^\circ$, is close to the present-day mean dynamical slope of Phobos and is adopted to reveal the underlying transport corridors, which may be representative either of past episodes of spontaneous regolith redistribution or of present-day transport initiated by external perturbations (e.g., seismic shaking, nearby impacts, fracture-related disturbances, thermal stresses).

This approach allows us to identify dynamically active regions and preferential transport corridors across the surface.}

\subsubsection{Regolith \del{M}\ins{m}igration \del{P}\ins{p}athways for the high-threshold end-member (30$^{\circ}$)}\label{Xsec7-5.1.1}
\label{friction_30}

\begin{figure*}

\centerline{\includegraphics[width=0.99\textwidth]{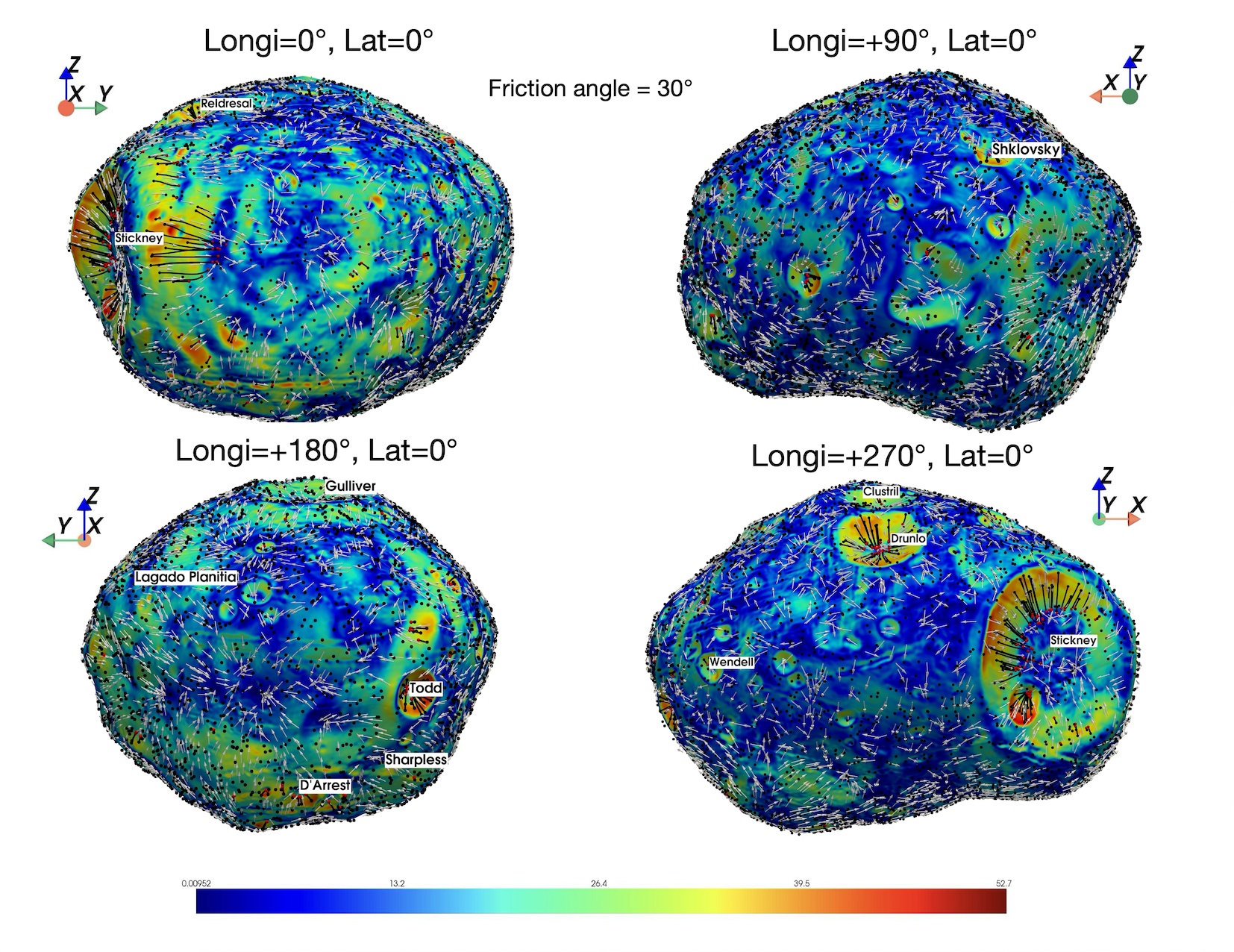}}
\caption{Surface trajectories simulated for $\phi = 30^{\circ}$, projected on Phobos' global topography (equatorial views).
Black lines highlight the Regolith Migration Pathways (RMPs), defined as preferred routes of regolith transport under the combined effects of self-gravity, centrifugal, and tidal accelerations.}
     \label{fig:appendix_trajectories_30_deg_2panels_poles}
 \end{figure*}

\begin{figure*}

\centerline{\includegraphics[width=0.95\textwidth]{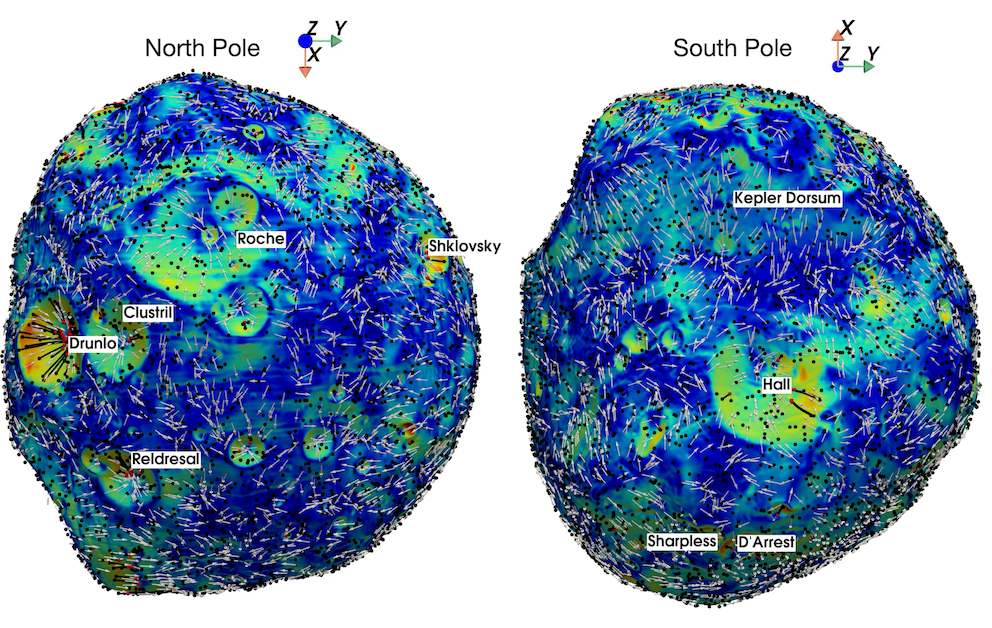}}
\caption{Surface trajectories simulated for $\phi = 30^{\circ}$, projected on Phobos' global topography (equatorial views).
Black lines highlight the Regolith Migration Pathways (RMPs), defined as preferred routes of regolith transport under the combined effects of self-gravity, centrifugal, and tidal accelerations. The polar regions are shown.
}     \label{fig:appendix_trajectories_30_deg_4panels}
 \end{figure*}

{ We explored a high-threshold end-member case by setting the friction angle to} $\phi_{static}=\phi_{dynamic}=30^\circ$ (\linkref[Figs.]{\ref{fig:appendix_trajectories_30_deg_2panels_poles}} and \ref{fig:appendix_trajectories_30_deg_4panels}). This choice is motivated by recent constraints on the mechanical properties of rubble-pile asteroids visited by spacecraft: \del{Robin {\rm et al.}} \cite{robin2024} derived bulk internal friction angles of $\sim 30^\circ$ (with Dimorphos at $32.7 \pm 2.5^\circ$) and reported comparable values for Itokawa, Ryugu, and Bennu, suggesting a common granular/boulder-scale mechanical behavior across these bodies. A friction angle in this range is also consistent with classical estimates for lunar regolith, where angles of internal friction are commonly reported in the $\sim 30^\circ$--$50^\circ$ interval depending on density and confining pressure \citep{mitchell1972}. { In our framework, adopting $\phi=30^\circ$ simultaneously as static and dynamic friction represents a conservative ``maximum-resistance'' scenario: any predicted motion then corresponds to trajectories that can be triggered spontaneously by the local dynamical slope alone, without invoking additional external perturbations.}

{ Under this stringent assumption, we find that: (i) downslope transport is limited to the interior of  the major craters, and also along the eastern outer slope of Stickney crater; (ii) no significant transport is found in most of the dynamical features identified in} \linkref[Section]{\ref{section:accel_dynamical_regions}}.

{ Concerning craters, the persistence of mobility under $\phi=30^\circ$ indicates that the present-day dynamical forcing and topographic configuration can still drive regolith motion there, continuously stripping the surface and favoring the exposure of deeper, less space-weathered material. This interpretation is consistent with the long-standing spectral dichotomy of Phobos, where the ``blue unit'' is concentrated around Stickney and is generally interpreted as fresher or more recently exposed material relative to the} globally dominant ``red unit'' \citep{thomas2011,fraeman2014,fraeman2012}. Moreover, dynamical models explicitly link the blue-unit distribution to ongoing surface refreshing by eccentricity-driven grain motion, which can excavate material faster than the space-weathering timescale in steep, highly varying-slope regions around Stickney \citep{Ballouz2019}. The fact that our $\phi=30^\circ$ case isolates only a small subset of mobile areas therefore supports the idea that these locations are among the most plausible candidates for present-day activity.

{ Importantly, areas that remain mobile even for $\phi=30^\circ$ are also the most plausible candidates for sustained dynamical activity under present-day or near-present-day topographic conditions. If local slopes were steeper before progressive mass wasting relaxed the surface, the same regions would likely have been even more favorable to transport. In that sense, the $\phi=30^\circ$ mobility patches may be regarded as robust candidates for preferential activity.}

{ Using such a high-threshold end-member (30$^{\circ}$) strongly restricts the expression of large-scale migration pathways within the present-day topographic framework. It leads to a model in which most of the surface remains inactive, with motion largely confined to the steepest terrains such as major crater walls and scarps. As a result, its diagnostic value for revealing the broader transport corridors that may have operated before or during resurfacing is limited. For this reason, in the following section we explore a lower-threshold end-member, close to the present-day mean dynamical slope of Phobos, in order to better reveal the potential Regolith Migration Pathways and their preferential directions.}

\subsubsection{Regolith \del{M}\ins{m}igration \del{P}\ins{p}athways for the low-threshold end-member (14$^{\circ}$)}\label{Xsec8-5.1.2}

{ We computed $\sim$5000 surface-constrained trajectories from randomly distributed starting points across the surface of Phobos, with motion initiated from rest ($v_0 = 0$) and using $\phi_{\text{static}} = \phi_{\text{dynamic}} = 14^{\circ}$. This setup is intended to reveal regions where Regolith Migration Pathways may have operated either during past episodes of spontaneous regolith redistribution, when local slopes were steeper than today, or under transport initiated by external perturbations.} Although this approach is simple, it provides a qualitative diagnostic of past potential surface motion. Reconstructing Phobos' paleomorphology would be preferable, but is currently infeasible due to the poorly constrained dynamical history.

\begin{figure*}

\centerline{\includegraphics[width=0.95\textwidth]{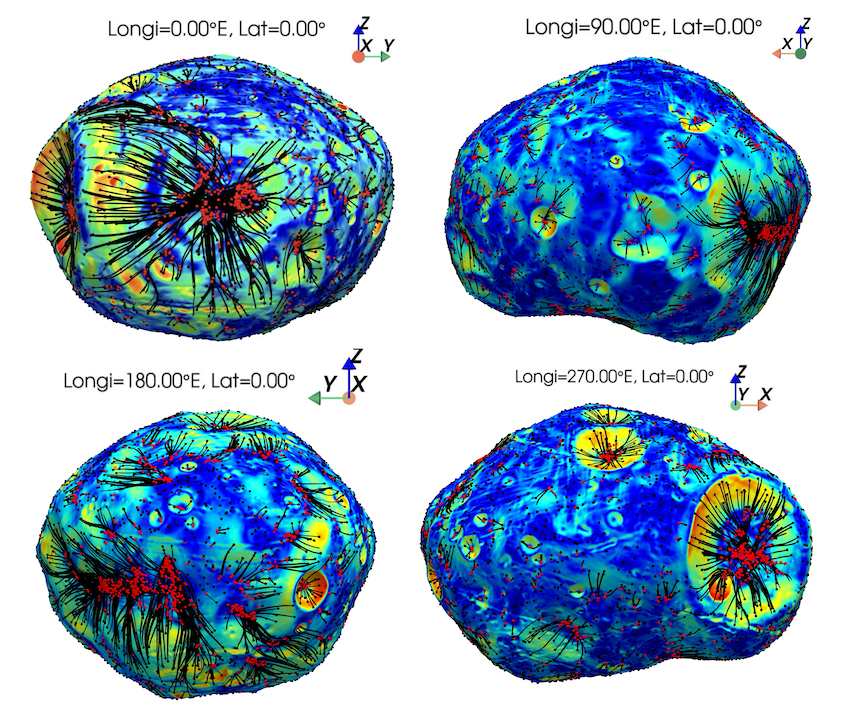}}
\caption{Surface trajectories simulated for $\phi = 14^{\circ}$, projected on Phobos' global topography (equatorial views).
Black lines highlight the RMPs, i.e. preferred routes of regolith transport, under the combined effects of self-gravity, centrifugal, and tidal accelerations. These pathways delineate regions of potential past or externally triggered surface mobility. Black dots designate starting points of the surface-constrained trajectories. Red dots show the end points of the computed pathways with non-zero lengths.
Major craters such as Stickney show converging downslope motions toward their floors, while discrete RMP networks connect large-scale morphological units, including the \textit{Eastern Outer Slope of Stickney Crater} (Feature~A) and the \textit{Sub-Mars Low} (Feature~B).
The correspondence between Regolith Migration Pathways and previously  dynamical regions identified on the basis of the acceleration maps (\linkref[\del{Figure}\ins{Fig.}]{\ref{figure:accel_map_6panels_FEATURES}}) supports the interpretation of long-term regolith transport shaping the surface of Phobos.
}
     \label{fig:trajectories_14_deg_4panels}
 \end{figure*}

\begin{figure*}

\centerline{\includegraphics[width=0.95\textwidth]{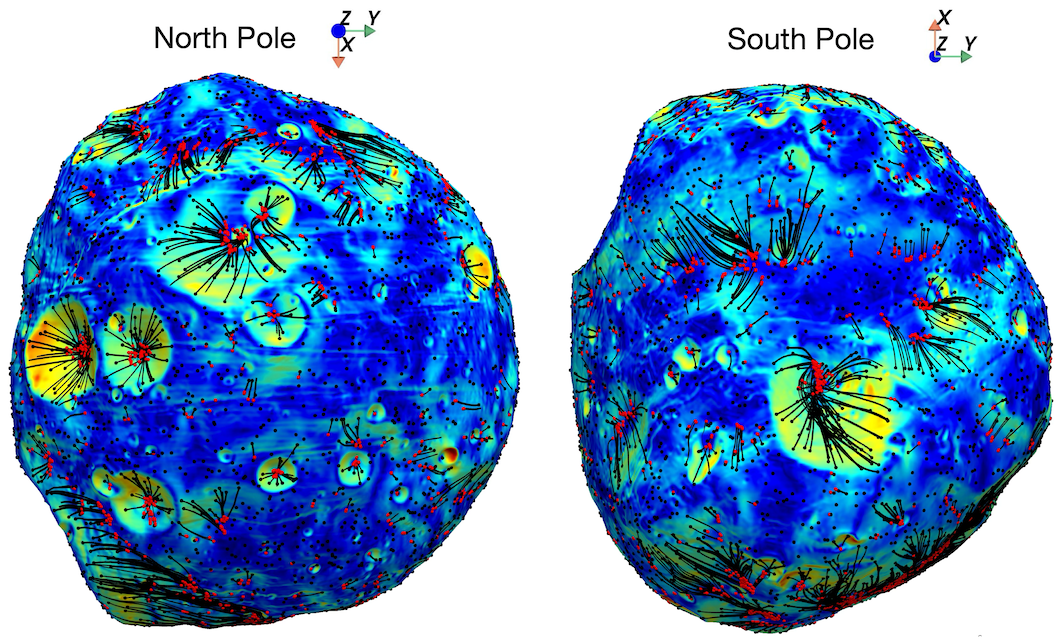}}
\caption{Simulated regolith trajectories at the north and south poles of Phobos for $\phi = 14^{\circ}$.
Black lines mark the Regolith Migration Pathways (RMPs).
At the north pole, limited RMP activity is observed except along \textit{Grildrig Ridge} (Feature~G), where material flows toward the anti-Mars side.
In contrast, the south polar region exhibits distinct pathways bordering \textit{South Pole Troughs} (Features~I and~H), suggesting localized regolith transport along scarps, while their interiors appear largely quiescent.
These results indicate that Phobos' polar regions may have experienced episodic mass wasting followed by the accumulation of fine-grained regolith, now forming relatively flat and stable surfaces. }
     \label{fig:trajectories_14_deg_poles}\vspace{3pt}
 \end{figure*}

Our results are shown in \linkref[\del{Figures}\ins{Figs.}]{\ref{fig:trajectories_14_deg_4panels}}-\ref{fig:trajectories_14_deg_poles}. { At first glance, most of the surface appears dynamically quiet, even for such a low-threshold end-member.} Major craters remain the most active regions, displaying downslope pathways directed toward their floors, as expected. A limited set of discrete zones shows potential surface mobility, some coinciding with the dynamical regions defined from the acceleration maps (see \linkref[Section]{\ref{section:accel_dynamical_regions}} and \linkref[\del{Figure}\ins{Fig.}]{\ref{figure:accel_map_6panels_FEATURES}}), and others deviating slightly. We provide in \linkref[Appendix]{\ref{subsec:slope_stability}} a detailed analysis of the distributions of trajectory lengths and durations. For $\phi = 14^{\circ}$, the median trajectory length is about 1.3\del{ }\ins{\thinspace }km (\linkref[\del{Figure}\ins{Fig.}]{\ref{fig:trajlength_cdf}}) and the median trajectory duration is about 0.68\del{ }\ins{~}h (\linkref[\del{Figure}\ins{Fig.}]{\ref{fig:trajtime_cdf}}), { both much shorter than Phobos' orbital period. These values should, however, be interpreted with caution, since this low-threshold case is used here primarily to reveal migration pathways and does not necessarily represent the actual frictional behaviour of the surficial material.}

The regions traversed by the black trajectory traces in \linkref[Fig.]{\ref{fig:trajectories_14_deg_4panels}}, and subsequent figures, delineate what we term Regolith Migration Pathways (RMPs).
Below, we describe these RMPs and compare them with the dynamical regions identified in \linkref[Section]{\ref{section:accel_dynamical_regions}}.

\begin{itemize}
\item \textbf{Sub-Mars side (longitude = 0$^{\circ}$, latitude = 0$^{\circ}$):}
The sub-Mars point lies at the convergence of numerous trajectories initiated on the eastern outer slope of Stickney crater and from mid-latitude terrains (up to $\sim$50$^{\circ}$S).
Most pathways terminate within a $\sim10^{\circ}$ window in both longitude and latitude, centered on the sub-Mars point.
These RMPs correspond closely to dynamical Features~A (\textit{Eastern Outer Slope of Stickney Crater}) and~B (\textit{Sub-Mars Low}).

\item \textbf{Trailing side (longitude = +90$^{\circ}$, latitude = 0$^{\circ}$):}
This region exhibits few RMPs, apart from those confined to major craters near the equator.
Feature~C (\textit{Anti-Mars Southern Troughs}), centered on the anti-Mars side, is visible in the eastern part.

\item \textbf{Anti-Mars side (longitude = +180$^{\circ}$, latitude = 0$^{\circ}$):}
This hemisphere is very active.
Large-scale migration pathways are visible within Feature~C (\textit{Anti-Mars Southern Troughs}), and  converge toward a longitudinally extended valley located just south of the equator, spanning nearly 40$^{\circ}$ in longitude.
Along the circular ridge identified as Feature~D (\textit{Far-Side Annular Ridge}), RMPs consistently follow downslope trajectories along the ridge flanks around its entire circumference.

\item \textbf{Leading side (longitude = +270$^{\circ}$, latitude = 0$^{\circ}$):}
Apart from Stickney crater, which channels regolith toward its center, and a few northern craters, this hemisphere shows little evidence of extensive RMPs.
Some pathways are visible at the southern margin of Feature~F (\textit{Leading Side Southern Low}), but the central area of Feature~F shows no motion, likely because it is mantled by a thick, nearly horizontal regolith layer that now masks the original relief. In such a depositional low, progressive infill would have reduced the local slopes to values below the considered threshold, so that the regolith remains stable even though the local acceleration field is non-zero as shown in \linkref[\del{Figure}\ins{Fig.}]{\ref{figure:accel_map_6panels_FEATURES}}d.

\item \textbf{Northern polar region (latitude = +90$^{\circ}$):}
Most of this region is dynamically stable, except along the ridge-like structure identified as Feature~G (\textit{Grildrig Ridge}), where RMPs direct material downslope toward the anti-Mars side, connecting with Feature~D (\textit{Far-Side Annular Ridge}). See \linkref[\del{Figure}\ins{Fig.}]{\ref{figure:accel_map_6panels_FEATURES}}e.

\item \textbf{Southern polar region (latitude = -90$^{\circ}$):}
This area encompasses Features~I and~H (\textit{South Pole Troughs}), which suggest potential regolith accumulation near the pole.
However, RMPs are restricted to the edges of these features, corresponding to local ridge or scarp structures, whereas their central portions appear largely immobile.
This pattern suggests that Features~I and~H are filled with a relatively flat, fine-grained regolith layer, which has become stabilized after progressive infilling under the acceleration field configuration shown in \linkref[\del{Figure}\ins{Fig.}]{\ref{figure:accel_map_6panels_FEATURES}}f.
\end{itemize}

\subsection{Comparison of Regolith Migration Pathways with Phobos \del{S}\ins{s}urface imagery}\label{Xsec9-5.2}

\label{section_comparison_RMP_with_images}

 { We compare the Regolith Migration Pathways with remote-sensing observations to assess whether regions of inferred regolith mobility display distinct morphological or spectral characteristics relative to dynamically stable areas. To this end, we use images from MRO/HiRISE and Mars Express/HRSC and compare them with the Regolith Migration Pathways computed for the low-threshold end-member ($14^{\circ}$). This comparison is not intended as a definitive validation of the model, nor as proof of a unique origin for the red and blue spectral units. Rather, it serves as a qualitative consistency test to evaluate whether the present-day dynamical field and topography are compatible with the large-scale distribution, exposure, reworking, or mixing of surface materials once emplaced at the surface.}

\subsubsection{Eastern \del{O}\ins{o}uter \del{S}\ins{s}lope of \del{S}\ins{s}tickney (A) and \del{S}\ins{s}ub-mars \del{L}\ins{l}ow (B)}\label{Xsec10-5.2.1}

\label{section_east_of_stickney_VS_images}

\begin{figure}
\centerline{\includegraphics[width=0.4\linewidth]{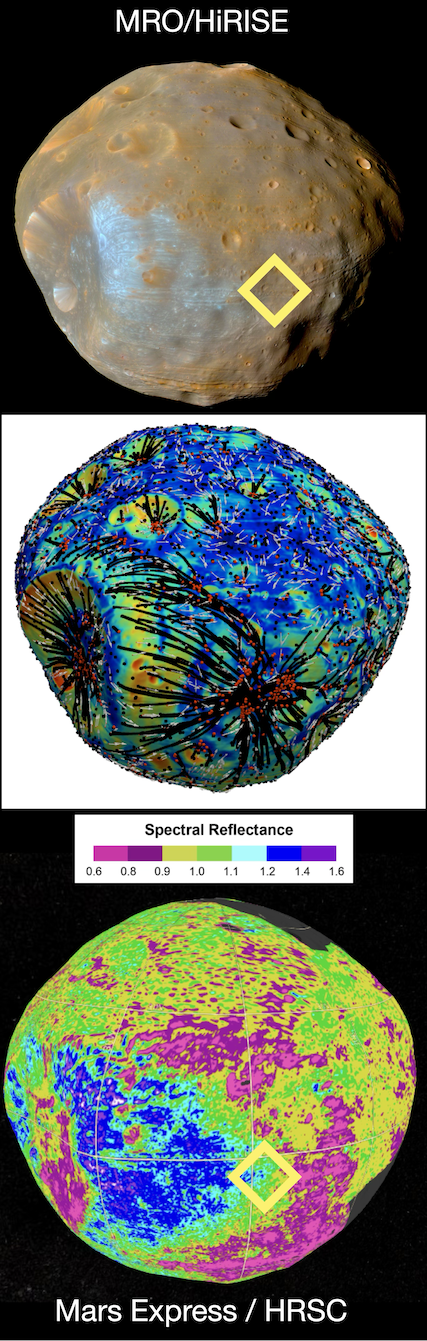}}
\caption{\textbf{Top:} Phobos nearside as observed by the MRO/HiRISE instrument in a three-color composite image \citep{thomas2011}.
\textbf{Middle:} 3D visualization of Phobos' dynamical surface. Black lines indicate Regolith Migration Pathways (RMPs) computed for $\phi = 14^{\circ}$; white arrows show acceleration vectors projected onto the local topography; and red dots mark the resting positions of regolith trajectories.
\textbf{Bottom:} Visible to near-infrared reflectance ratio (G+B)/(2\,NIR), obtained with the Mars Express HRSC instrument, highlighting spectrally ``blue'' regions near Stickney crater. This projection was generated using the NASA ``Phobos Trek'' multi-instrument mapping tool \citep{Karachevtseva_2014_Phobos_Trek} (\url{https://trek.nasa.gov/phobos/}). The yellow hexagon marks the area where RMPs terminate and red dots are abundant in the middle panel; this location corresponds to the sub-Mars point of Phobos.
    }
    \label{fig:phobos_sub_MARS_MRO_HIRIZE_comparison}
\end{figure}

At the sub-Mars point, spectrally ``blue'' regions have long been associated with the outer steep slopes of the Stickney crater \citep{thomas2011}.
\linkref[\del{Figure}\ins{Fig.}]{\ref{fig:phobos_sub_MARS_MRO_HIRIZE_comparison}} compares visible images from the MRO/HiRISE instrument with our derived Regolith Migration Pathways.
The RMPs follow remarkably well the triangular structure of the large blue unit located east of Stickney (compare the middle panel of \linkref[\del{Figure}\ins{Fig.}]{\ref{fig:phobos_sub_MARS_MRO_HIRIZE_comparison}} with the top and bottom panels), suggesting that the blue unit may correspond to active areas of regolith transport. The bottom image of \linkref[\del{Figure}\ins{Fig.}]{\ref{fig:phobos_sub_MARS_MRO_HIRIZE_comparison}}  shows a spectral reflectance index calculated using Mars Express HRSC instrument. This spectral index is a ratio of the albedo in three band filters: (G+B)/(2 NIR).
The endpoints of the pathways (red dots in the middle panel) converge near the sub-Mars point (yellow diamond), within a spectrally ``green'' crater (spectral slope $\sim 1$) outlined by a yellow hexagon. In contrast, zones of possible regolith transport where material does not settle appear spectrally blue, potentially exposing fresher, less weathered subsurface material. Conversely, areas north of the sub-Mars point where the regolith appears largely immobile display a more reddish coloration. This pattern may indicate that depositional areas which have become dynamically inactive gradually evolve from a spectrally neutral (green) appearance toward a redder tone over time.

\begin{figure*}

\centerline{\includegraphics[width=0.95\textwidth]{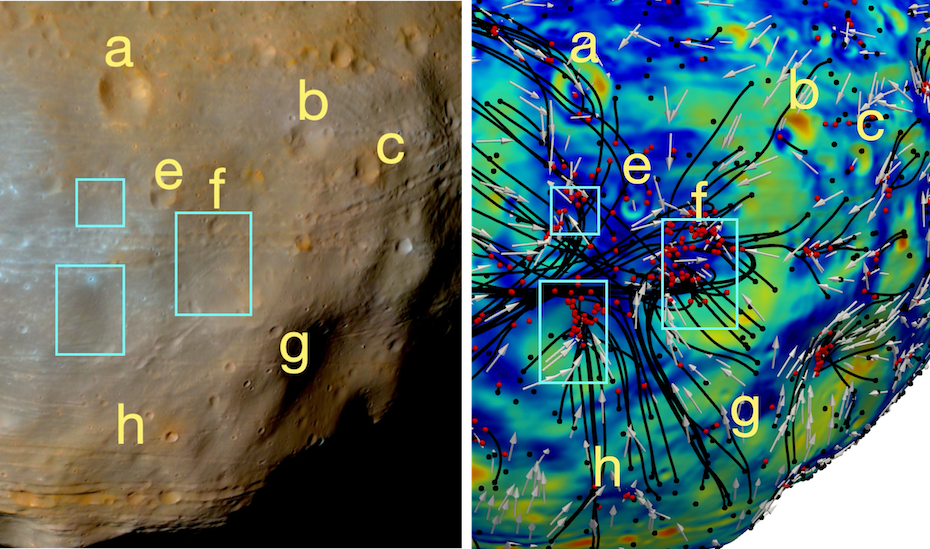}}
\caption{ Zoom on the sub-Mars low region. Letters a to h mark topographic features used as control points to aid interpretation of the images. Blue rectangles highlight potential regions of regolith accumulation (corresponding to clusters of red endpoints of the Regolith Migration Pathways) and show smoother surfaces, whereas the area extending between control points \textit{a} and \textit{b} shows no regolith motion (right panel) and appears rougher and more densely populated with small craters (left panel).
\textbf{Left:} Detail of the MRO/HiRISE image (top of \linkref[\del{Figure}\ins{Fig.}]{\ref{fig:phobos_sub_MARS_MRO_HIRIZE_comparison}}) of the sub-Mars region.
\textbf{Right:} Dynamical map where black lines show the Regolith Migration Pathways and red dots mark the endpoints of the trajectories.
   }
    \label{fig:sub_mars_zoom_control_points}
\end{figure*}

A zoom on the sub-Mars point is shown in \linkref[\del{Figure}\ins{Fig.}]{\ref{fig:sub_mars_zoom_control_points}}. The left panel displays a detail of the HiRISE image shown at the top of \linkref[\del{Figure}\ins{Fig.}]{\ref{fig:phobos_sub_MARS_MRO_HIRIZE_comparison}}. Geographic control points labeled with letters a to h are used to identify the corresponding regions in the dynamical map on the right. In the right panel of \linkref[\del{Figure}\ins{Fig.}]{\ref{fig:sub_mars_zoom_control_points}}, the dynamical map is shown, including both the acceleration field and the Regolith Migration Pathways (RMPs). Blue rectangles highlight three clusters of RMP endpoints around control point \textit{e}. { Inspection of the corresponding areas in the HiRISE color image (left panel) shows that these regions exhibit comparatively smooth (i.e. mantled-looking) terrain, a more neutral (grayish) color, and a lower abundance of very small craters, consistent with surficial regolith accumulation.} In contrast, the area extending from control points \textit{a} to \textit{b}, where no regolith motion is predicted, appears rougher and more densely populated with small craters. This contrast, i.e. smooth, lightly cratered surfaces at RMP termini versus rough, heavily cratered terrains in dynamically inactive areas, is observed at several locations across the surface of Phobos.

\subsubsection{Northern \del{R}\ins{r}egions and \del{F}\ins{f}ar-side \del{A}\ins{a}nnular \del{R}\ins{r}idge (D)}\label{Xsec11-5.2.2}

\begin{figure*}

\centerline{\includegraphics[width=0.8\textwidth]{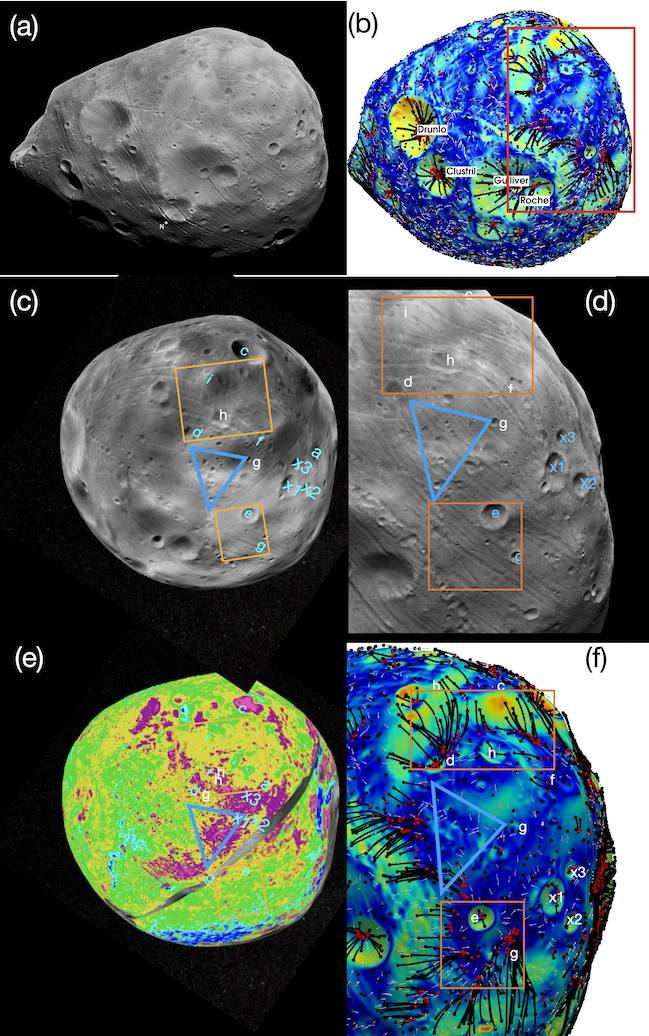}}
\caption{Comparison of Mars Express/HRSC image and spectral slope of the northern polar region of Phobos with dynamical map. (a) Mars Express/HRSC image of the polar regions. The north pole is marked by a white dot. (b) Dynamical map of the same area, with black lines showing Regolith Migration Pathways ($\phi = 14^{\circ}$) and red dots marking the endpoints of these trajectories. The red rectangle highlights a region of particular interest that is examined in more detail in panels c,d,e,f where the letters correspond to topographical control points.(c)  Mapping with Viking images obtained with the  NASA "Phobos Trek" multi-instrument mapping tool \citep{Karachevtseva_2014_Phobos_Trek} (\url{https://trek.nasa.gov/phobos/}). The  orange rectangles correspond to two remarkable regions with potential accumulation of regolith. The blue triangle corresponds to a region where material is static. (d) Zoom on the red-rectangular region displayed in panel b (e) Visible-to-near-infrared reflectance ratio obtained with Mars Express HRSC instrument (f) Dynamical map of the region displayed in panel d.}
    \label{fig:north_6_panels}
\end{figure*}

The northern region displays a heterogeneous surface (\linkref[\del{Figure}\ins{Fig.}]{\ref{fig:north_6_panels}}a and d).
The dynamical map (\linkref[\del{Figure}\ins{Fig.}]{\ref{fig:north_6_panels}}b) shows that most of this area is stable, with significant regolith motion limited to large craters and to a prominent circular structure corresponding to the northern part of Dynamical Feature~D, the \textit{Far-side Annular Ridge} (\linkref[Section]{\ref{section:accel_dynamical_regions}}).
Along this ridge, where Regolith Migration Pathways are present, the surface appears systematically smoother than the ridge interior, which lacks RMPs and remains rough and heavily cratered.

A more detailed comparison is shown in \linkref[\del{Figure}\ins{Fig.}]{\ref{fig:north_6_panels}}c to f, combining HRSC imagery (panels~c, d), a spectral slope map (panel~e), and the dynamical map (panel~f).
Clusters of RMP endpoints (orange rectangles in panel~d) coincide with smoother terrains in the HRSC images and with spectrally neutral (green) regions, similar to those observed near the sub-Mars point.
In contrast, areas with little or no predicted regolith motion (e.g., near point~\textit{f}) appear rougher and spectrally redder.
Together, these observations support the interpretation that dynamically active zones favor regolith accumulation, producing smooth, spectrally neutral surfaces, whereas dynamically stable terrains evolve toward rougher, redder states.

\subsubsection{Anti-mars \del{S}\ins{s}outhern \del{D}\ins{d}ynamical \del{T}\ins{t}roughs (C and E)}\label{Xsec12-5.2.3}

\label{section_discussion_southern_through_comparison_with_images}

\begin{figure*}

\centerline{\includegraphics[width=0.95\textwidth]{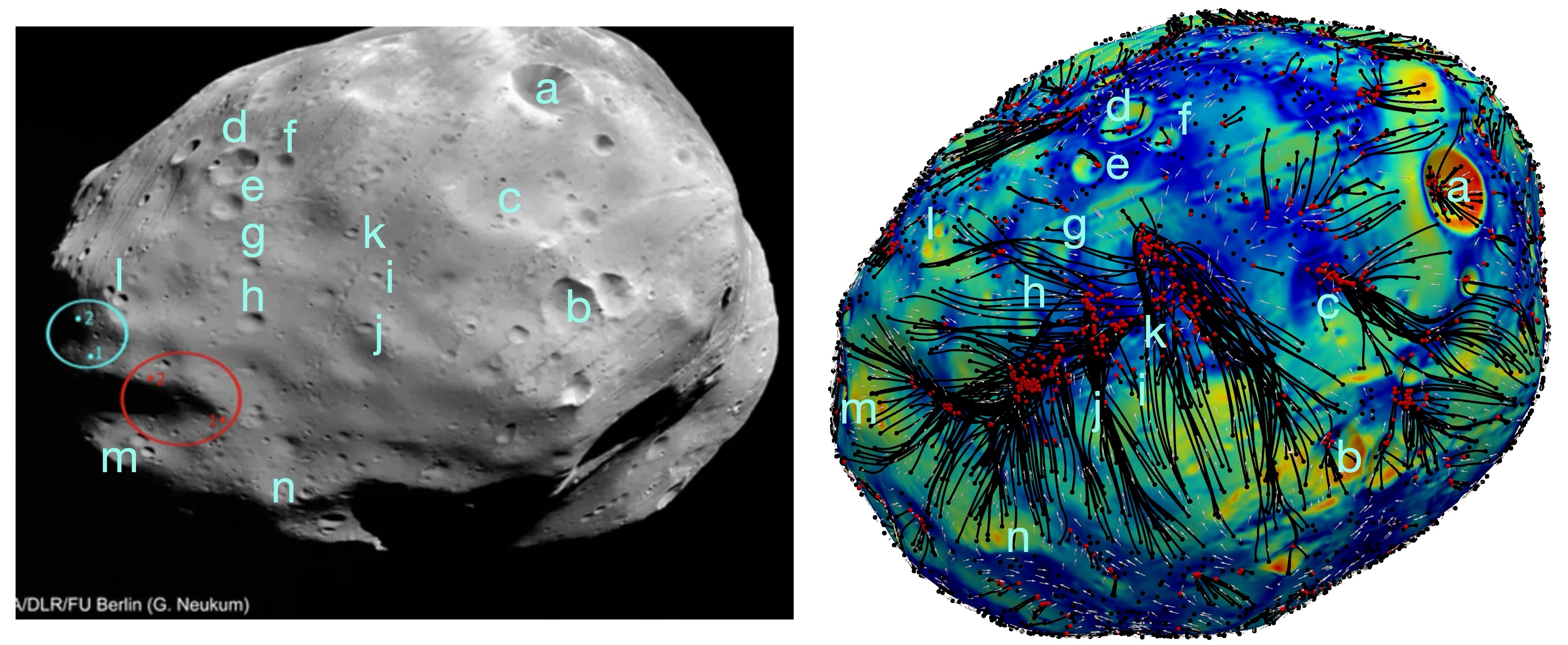}}
\caption{ These panels present a detailed analysis of the anti-Mars region. Letters correspond to topographic control points shown in both panels to facilitate comparison. \textbf{Left}: HRSC high-resolution image of the anti-Mars region. \textbf{Right}: Dynamical map of the region displayed in the left panel. Blue and red circles on the lower left corner of the left panel show the possible landing site of the former failed Phobos Grunt mission. }
    \label{fig:anti-mars_2@@panels_comparison}\vspace*{3pt}
\end{figure*}

The anti-Mars hemisphere, along with a portion of the southern terrain, is shown in \linkref[\del{Figure}\ins{Fig.}]{\ref{fig:anti-mars_2@@panels_comparison}}.
The anti-Mars side hosts an extensive region dense in Regolith Migration Pathways that converge along a band near the equator, culminating in a concentration of trajectory endpoints (red dots) between topographic points \textit{m}, \textit{j}, \textit{k}, and \textit{l}, and correspond to dynamical features C and E:  \textit{Anti-Mars Southern Troughs}.  The HRSC image (left panel) confirms that this region is remarkably smooth,
consistent with resurfacing by mobile regolith, possibly indicative of relatively recent activity.
These terrains are characterized by a low abundance of small craters and a coherent, mantled texture,
in  contrast with nearby rougher and more heavily cratered areas (e.g., around control points \textit{l}, \textit{m}, and \textit{n}).
Even craters located near points \textit{g}, \textit{h}, \textit{i}, \textit{j}, and \textit{k} appear partially infilled,
and linear grooves progressively fade in the vicinity of point~\textit{j},
a pattern consistent with burial by deposited regolith.

The spatial distribution of these smooth terrains matches the RMPs and their termination zones
displayed (\linkref[\del{Figure}\ins{Fig.}]{\ref{fig:anti-mars_2@@panels_comparison}} right).
According to the dynamical map, the mobilized material likely originates from the surrounding highlands south of the equator,
near control points \textit{i}, \textit{m}, \textit{n}, and \textit{g}.
These areas are topographically elevated and exhibit rough surface textures,
making them natural candidate source regions for downslope regolith transport.

The correspondence between smooth, mantled-looking terrains and Regolith Migration Pathways is further strengthened
when compared with dynamical simulations that allow trajectories to temporarily detach from the surface
(\linkref[Appendix]{\ref{appendix:flying_particles_dynamics}}).
\linkref[\del{Figure}\ins{Fig.}]{\ref{fig:appendix_trajectories_flying_part_14_deg_4panels}}c illustrates that only minor differences arise
between surface-constrained (right panel of \linkref[\del{Figure}\ins{Fig.}]{\ref{fig:anti-mars_2@@panels_comparison}})
and unconstrained simulations.
These differences are highlighted by the orange circle in
\linkref[\del{Figure}\ins{Fig.}]{\ref{fig:appendix_trajectories_flying_part_14_deg_4panels}},
corresponding to the region extending between control points \textit{c} and \textit{k}.

In this area, the surface appears very smooth, with several craters partially infilled.
The unconstrained simulations (\linkref[\del{Figure}\ins{Fig.}]{\ref{fig:appendix_trajectories_flying_part_14_deg_4panels}}c) indicate that trajectories originating southeast of control point~\textit{c}
can enter the crater near \textit{c}, undergo short ballistic excursions over its rim,
and subsequently settle in the region between points \textit{c} and \textit{k}.
This mechanism provides a plausible explanation for the observed infilling and reinforces the interpretation
that localized regolith transport involving brief detachment events may contribute to surface smoothing,
while leaving the large-scale regolith redistribution pattern largely unchanged.

These observations strengthen the link between surface smoothness and regolith mobility: smooth terrains (points \textit{e}, \textit{f}, \textit{k}, \textit{i}, \textit{g} in \linkref[Fig.]{\ref{fig:anti-mars_2@@panels_comparison}}) likely correspond to depositional zones along RMPs, whereas rough, elevated areas (points \textit{l}, \textit{m}, \textit{n}, \textit{a}, \textit{c} in \linkref[Fig.]{\ref{fig:anti-mars_2@@panels_comparison}}) act as source regions for downslope transport.
No high-resolution spectral data are currently available for this area.

\subsubsection{South \del{P}\ins{p}ole \del{T}\ins{t}roughs (H and I)}\label{Xsec13-5.2.4}

\label{section_discussion_south_pole_comparison_with_images}

\begin{figure*}

\centerline{\includegraphics[width=0.95\textwidth]{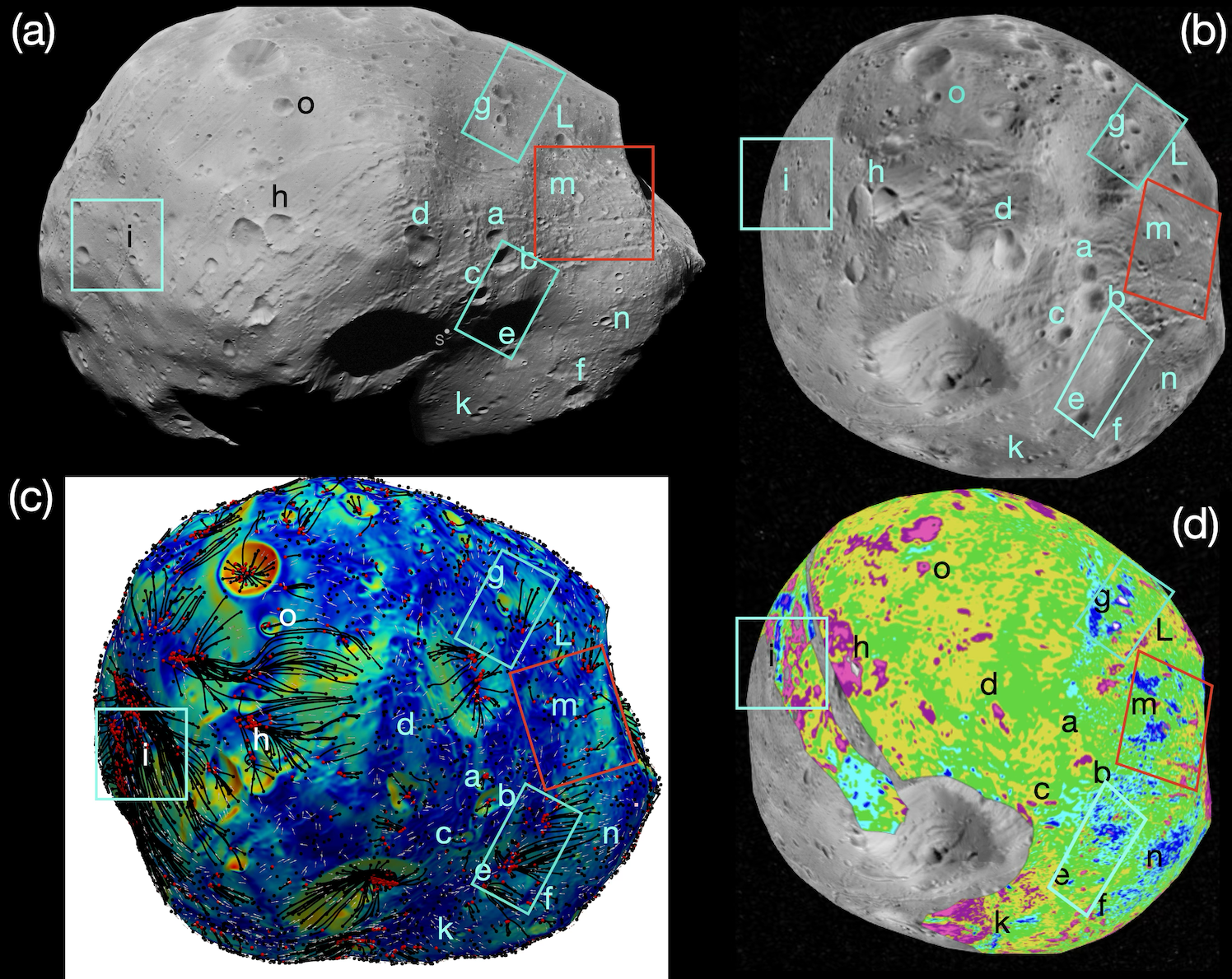}}
\caption{ This set of images presents a detailed analysis of the southern regions of Phobos (near point \textit{k}) and the anti-Mars side (near point \textit{i}). Letters correspond to topographic control points reported in the four panels, to facilitate comparison despite the changing viewing geometry. (a) Mars Express/HRSC view of the south polar regions.
(b) Mapping with Viking images obtained using the NASA ``Phobos Trek'' multi-instrument mapping tool \citep{Karachevtseva_2014_Phobos_Trek} (\url{https://trek.nasa.gov/phobos/}).
(c) Dynamical map of the region shown in panel (a).
(d) Visible-to-near-infrared reflectance ratio obtained with the Mars Express HRSC instrument.
}
    \label{fig:south_4_panels_comparison}
\end{figure*}

We now focus on the southern regions of Phobos, illustrated in \linkref[\del{Figure}\ins{Fig.}]{\ref{fig:south_4_panels_comparison}}.
A wide variety of terrains can be observed.
The blue rectangles mark areas of potential regolith accumulation, as indicated by the dynamical map (panel~c) computed for $\phi = 14^{\circ}$.
The area surrounding the topographic points \textit{e}, \textit{f} and \textit{k} corresponds to the dynamical Feature~I (\textit{low South Pole}).
While this region appears remarkably smooth in HRSC images (panel~a), only a few localized zones seem to coincide with the endpoints of Regolith Migration Pathways, such as the area enclosed by the rectangle between points \textit{b}, \textit{e}, and \textit{f}.
These regions appear either spectrally neutral or slightly blue in panel~(d), consistent with relatively recent regolith transport.

The surface texture near point~\textit{k} is extremely smooth and could alternatively be interpreted as a relatively recent ejecta blanket, possibly originated near point~\textit{e}.
In contrast, elevated terrains with higher slopes, such as those around topographic points \textit{m} and \textit{g}, appear systematically rougher and exhibit strong blue spectral indices, compatible with recently denuded areas.
Thus, while smoother terrains tend to show neutral or weakly blue colors, similar to the northern regions near Feature~D (\textit{Far-Side Annular Ridge}), the southern terrains display a greater diversity in spectral slopes, perhaps reflecting differences in regolith maturity or resurfacing history. Interestingly, rough terrains that appear dynamically less active in the surface-constrained model (e.g., near points \textit{g} and \textit{m}) exhibit a pronounced blue spectral signature, suggesting limited regolith mixing and the possible exposure of less weathered material. To better interpret these areas, it is necessary to analyse the results of the unconstrained simulations presented in~\linkref[Appendix]{\ref{appendix:flying_particles_dynamics}}. \linkref[\del{Figure}\ins{Fig.}]{\ref{fig:appendix_trajectories_flying_part_14_deg_north_south}}-right (orange ellipse) shows that the area surrounding control points \textit{g} and \textit{m} is in fact dynamically active, giving rise to regolith migration and consequently to the denudation of the terrains that exhibit blue spectra.

\section{Discussion and \del{C}\ins{c}onclusion}\label{Xsec14-6} \label{Conclusion}

We have presented a novel approach to study the mobility of surface material on small bodies  and have applied it to Phobos.

This study reveals a set of spatially coherent \textbf{Surface Dynamical Features}, defined as regions where the surface acceleration field exhibits a consistent structure (Features~A--I; \linkref[Section]{\ref{section:accel_dynamical_regions}}). Among these, two major dynamical systems stand out: regions near the sub-Mars point (Features~A and~B, \textit{Eastern outer slope of Stickney crater} and \textit{Sub-Mars Low}), which act as collection zones for material sourced from Stickney crater, and equatorial trough-like regions on the anti-Mars hemisphere (Features~C and~E, \textit{Anti-Mars Southern Dynamical Troughs}), which are favorably located to accumulate material from both hemispheres.

Then, we derived potential \textit{Regolith Migration Pathways} (RMPs) that describe the preferred routes of surface material transport across Phobos. { For a high-threshold end-member of about $30^{\circ}$, consistent with some friction estimates for small bodies} \citep{robin2024}{ , mobility is largely restricted to steep crater walls and does not correlate with the main observed surface characteristics. In contrast, for a low-threshold end-member comparable to Phobos' mean dynamical slope ($\sim 14^{\circ}$), the model reveals an interconnected network of migration pathways spanning much of the surface. While this low-threshold end-member is not intended to represent present-day spontaneous motion, it provides a diagnostic view of the large-scale Regolith Migration Pathways that may have operated under past or transient conditions, and shows good qualitative agreement with many observed surface features.}

Together, these pathways delineate dynamically active regions where long-term regolith motion can contribute to surface reshaping, crater infilling, and the development of spectral and morphological heterogeneity.

\begin{itemize}
    \item \textbf{Sub-Mars hemisphere:}
    The triangular-shaped region east of Stickney crater (Features~A, \textit{Eastern Outer Slope of Stickney Crater} and~B \textit{Sub-Mars Low}) has long been known to host spectrally blue material.
    Our simulations show that Regolith Migration Pathways for $\phi = 14^{\circ}$ reproduce this geometry remarkably well, suggesting that this area records a large-scale, long-lasting episode of mass movement.
    Most trajectories originating near the Stickney crater outer slope terminate close to the sub-Mars point, consistent with the accumulation of mobile regolith in this region.
    Spectrally, the sub-Mars point itself appears neutral, while the adjacent steep slopes, where regolith does not accumulate, are distinctly blue.
    This pattern supports a time-evolution of spectral slopes, where mobile, freshly exposed material appears blue and stable surfaces evolve from green (i.e. neutral) toward  reddish tones (\linkref[\del{Figures}\ins{Figs.}]{\ref{fig:phobos_sub_MARS_MRO_HIRIZE_comparison}}-\ref{fig:sub_mars_zoom_control_points}).

    \item \textbf{Northern hemisphere:}
    Apart from local motion within craters, evidence of downslope transport is visible along Feature~D (\textit{Far-Side Annular Ridge}), primarily on the far side of Phobos.
    Near the pole, HRSC images reveal a mosaic of spectrally neutral and red terrains corresponding, respectively, to smooth accumulative surfaces and rough, static ones.
    This qualitative correspondence between regolith accumulation, smoothness, and spectral neutrality strengthens the link between surface texture and dynamical activity (\linkref[\del{Figure}\ins{Fig.}]{\ref{fig:north_6_panels}}c to f).

    \item \textbf{Southern hemisphere:}
    Near the south pole, a broad dynamical depression comprising Features~H and~I (\textit{South Pole Troughs}) is observed close to the Kepler Dorsum region.
    HRSC images show that this area is exceptionally smooth, yet the local slopes are so gentle that accumulation occurs primarily along the edges of the troughs (\linkref[\del{Figure}\ins{Fig.}]{\ref{fig:south_4_panels_comparison}}).
    While smooth regions correlate well with low-slope dynamical features, the spectral correlation is weaker, possibly due to differing surface ages or the presence of an ejecta blanket from the nearby Hall crater.

    \item \textbf{Anti-Mars hemisphere:}
    One of the most striking dynamical regions corresponds to Features~C and~E (\textit{Anti-Mars Southern Troughs}), which show dense clustering of RMP endpoints (\linkref[\del{Figure}\ins{Fig.}]{\ref{fig:anti-mars_2@@panels_comparison}}).
    This area aligns with extremely smooth terrains in HRSC imagery, indicative of regolith accumulation and resurfacing, although no high-resolution spectral data currently exist for this region.
\end{itemize}

Overall, our results indicate that Phobos' present-day surface reflects the coupled effects of topography, gravity-driven regolith transport, and tidal forcing. Dynamical convergence zones act as regolith sinks, whereas elevated or rough terrains serve as source regions.

\begin{figure*}

\centerline{\includegraphics[width=0.95\textwidth]{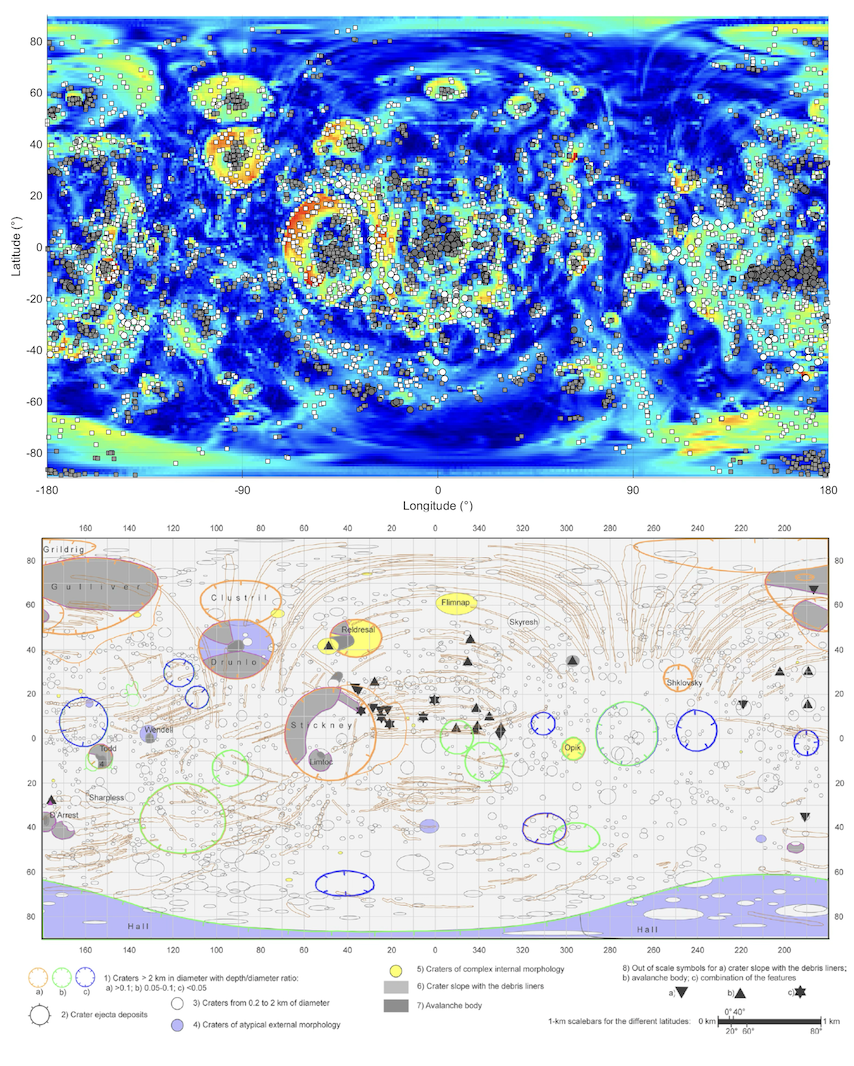}}
\caption{Comparison between modeled regolith accumulation zones and observed mass-wasting features on Phobos. Top panel: Global map of Phobos showing the local dynamical slope (color scale) with superimposed endpoints of Regolith Migration Pathways (gray squares and circles), representing predicted accumulation zones for mobile regolith. White squares and circles show the starting points of Regolith Migration Pathways. For both start and end points, squares denote short pathways, whereas circles denote pathways lasting longer than 2~h. Bottom panel: Geological feature map of Phobos adapted from \del{Basilevsky {\rm et al.} (2014)} \cite{Basilevsky_2014}, highlighting mapped landslides, avalanche bodies, and mass-wasting features (black triangles and stars). Despite slight differences in map projection and plotting conventions, regions of high endpoint density in the dynamical model broadly coincide with independently identified landslide locations, supporting a dynamical control on regolith redistribution across the surface of Phobos.}
     \label{fig:comparison_geological_features_and_endpoints}
 \end{figure*}

A comparison between the dynamical map and the geological feature map shown in \linkref[\del{Figure}\ins{Fig.}]{\ref{fig:comparison_geological_features_and_endpoints}} reveals a strong qualitative correspondence between modeled regolith accumulation zones  (using the low end-member $\phi = 14^{\circ}$) and independently identified mass-wasting features on Phobos (black triangles and stars, lower panel) reported in the work of \del{Basilevsky {\rm et al.} (2014)} \cite{Basilevsky_2014}. The endpoints of the Regolith Migration Pathways (gray squares and circles in the upper panel, with $\phi = 14^{\circ}$) cluster preferentially in regions where landslides and avalanche bodies have been visually identified at the surface of Phobos (black triangles and stars in the lower panel of \linkref[\del{Figure}\ins{Fig.}]{\ref{fig:comparison_geological_features_and_endpoints}}).{ Note that minor differences may arise from the use of different map projections and plotting conventions. The Basilevsky map (bottom) is presented using a cylindrical projection, for which the reference radius is not specified. We therefore adopt a standard longitude-latitude representation in spherical coordinates. The differences between the two projections are minor but do not prevent clear identification of the different mass-wasting features.}

{ This study also suggests that some smooth-textured terrains may reflect past episodes of enhanced regolith mobility, potentially associated with locally steeper slopes that were later reduced by progressive relaxation. However, we do not regard this as a unique interpretation, since smooth terrains may also result from repeated regolith redistribution, impact-induced shaking, or other resurfacing processes.}

{ A useful independent test would be to compare the predicted regions of regolith accumulation with crater-density maps of Phobos, in order to assess whether zones of enhanced accumulation correspond to lower crater densities, and conversely whether more static regions are associated with higher crater densities. However, although such maps have been reported in the literature (e.g.,} \citealp{Salamuniccar_2014}){ , they do not appear to be readily accessible or downloadable at present.}

{ Most of the correlations identified in this study between the modeled Regolith Migration Pathways, smooth terrains, and red--blue spectral units are reproduced only for simulations using $\phi = 14^{\circ}$, a value significantly lower than friction angles inferred from boulder morphology on other small bodies} \citep{robin2024}{ , and close to the mean dynamical surface slope observed on Phobos }(\linkref[\del{Figure}\ins{Fig.}]{\ref{fig:slope_histrogram}}). { In contrast, simulations performed with a friction angle of $30^{\circ}$, consistent with more conservative estimates for granular materials on small bodies }\citep{robin2024}, predict very limited mobility outside steep crater walls and show no systematic correspondence with smooth terrains or spectral units (\linkref[Section]{\ref{friction_30}}).

This result should not be interpreted as evidence that $14^{\circ}$ represents the internal friction angle of Phobos' regolith, nor as a direct estimate of its present day frictional properties. Rather, in our simplified kinematic model, the $14^{\circ}$ case is used as a low threshold end member to reveal the broader organization of Regolith Migration Pathways under conditions of enhanced mobility, whether associated with steeper local slopes in the past or with transport triggered by external perturbations. In that sense, the $14^{\circ}$ case provides a pathway revealing reference level within the present day topography, not a direct physical measurement of regolith friction.

In addition, the use of the low end member, $14^{\circ}$, can be discussed in the context of the broader literature on long runout landslides. In such events, the apparent friction angle inferred at the scale of the whole deposit can be much lower than the internal friction angle of the material (see e.g. \citealp{Johnson_2016_low_friction_angle, Lucas_2024, Lei_2025}). This apparent friction is commonly expressed through the Heim ratio \citep{Lucas_2024}, defined as the ratio between the vertical drop and the horizontal runout distance. Low Heim ratios imply long runout and low apparent friction. Terrestrial and planetary examples show that apparent friction can decrease sharply with increasing runout length and with decreasing surface acceleration \citep{Lucas_2024}. Therefore, the apparent friction angle inferred from long runout deposits, or the effective friction angle used in a large scale transport model, should not be identified directly with the internal friction angle of the material \citep{Lei_2025}.  We do not claim that the same physical mechanisms responsible for long runout landslides operate on Phobos. Rather, the analogy is that macroscopic transport can be governed by an apparent or effective resistance that is lower than the internal friction angle, maybe due to self-organization processes not considered here (like acoustic fluidization as argued in \citealp{Johnson_2016_low_friction_angle}). This supports the idea that values lower than $30^{\circ}$ cannot be excluded a priori for large scale regolith transport, especially in weak gravity and for a highly porous surface layer.

The spatial agreement obtained for the $14^{\circ}$ case suggests that the present day acceleration field and topography may still retain the imprint of past or transient regolith redistribution processes. The modeled Regolith Migration Pathways therefore offer a useful basis for interpreting the large scale organization of surface transport on Phobos, and may help place future MMX samples in their broader geomorphological context.

\subsection{Possible \del{I}\ins{i}nterpretation of \del{R}\ins{r}ed, \del{N}\ins{n}eutral and \del{B}\ins{b}lue \del{U}\ins{u}nits}\label{Xsec15-6.1}

The model presented here provides a physical framework to interpret the spectral heterogeneity observed on Phobos' surface. In particular, the RMPs computed for  14$^\circ$ show systematic correlations with three main spectral classes: spectrally blue units, spectrally neutral (greenish) terrains, and spectrally red regions.

At the sub-Mars point, and in the northern and anti-Mars sectors, spectrally neutral, smooth terrains often coincide with clusters of RMP endpoints and with dynamically low-lying areas where downslope trajectories converge. These regions show a relative paucity of small craters and a mantled appearance, consistent with regolith accumulation and local resurfacing by material transported along RMPs. In this interpretation, the neutral (greenish) spectral signature reflects relatively young depositional surfaces that have been repeatedly reworked and smoothed by regolith infill, but not yet strongly reddened by space weathering.

By contrast, spectrally blue units are typically associated either with steep, dynamically active slopes where material cannot settle (e.g., flanks of Stickney and other craters), or with rough, high-standing terrains that appear denuded in the dynamical maps. These areas often show limited net deposition, and our simulations indicate that they can act as source regions feeding downslope transport. The blue spectral slope is therefore consistent with the exposure of relatively fresh material, either because fine regolith is being removed faster than it can accumulate, or because impacts and seismic shaking repeatedly strip or overturn the surface.

Finally, spectrally redder regions are most commonly found in dynamically quiet areas where the model predicts negligible regolith motion and where no significant RMP convergence is observed. These terrains tend to be rougher and more densely cratered, suggesting older, dynamically inactive surfaces that have not experienced substantial recent burial or mixing. In this picture, red units represent long-lived, slowly evolving surfaces, while neutral units mark younger depositional mantles, and blue units trace actively reworked or recently denuded material. This view is close to the Case 1 model discussed in \del{Basilevsky {\rm et al.} (2014)} \cite{Basilevsky_2014}.

Taken together, the good spatial { correlation} between RMPs, topographic smoothness, and spectral classes { suggests} that the present-day color pattern on Phobos { may} encode a dynamical stratigraphy: blue and neutral terrains { may} reflect zones of recent or ongoing regolith redistribution, whereas red terrains { may} correspond to regions that have remained dynamically stable over longer timescales.

\subsection{Implications for the MMX sampling sites}\label{Xsec16-6.2}
The forthcoming \textit{MMX} mission will return samples from two key regions of Phobos, near the sub-Mars and anti-Mars points, which our dynamical analysis identifies as distinct surface environments with contrasting regolith histories. The sub-Mars site lies within a major zone of material convergence (Features~A and~B), where regolith derived from the eastern outer slope of Stickney crater and adjacent terrains accumulates. Samples collected there are therefore expected to consist of well-mixed, mature regolith, integrating material from multiple source regions and recording a long history of surface processing and space weathering. In contrast, the anti-Mars site overlaps the \textit{Anti-Mars Southern Dynamical Troughs} (Features~C and~E), a smoother but dynamically active region where regolith motion may still be ongoing. Samples from this area may thus include younger, less-weathered material, potentially preserving information about Phobos' more pristine surface or shallow subsurface. Comparing samples from these two sites will provide a key test of our model predictions and help disentangle the respective roles of mechanical transport and compositional heterogeneity in Phobos' geological evolution.

Beyond sample provenance, the identification of dynamically quiet regions and preferential transport corridors is also relevant for surface operations. A transport corridor may be inactive at the time of landing, but it still marks the route along which regolith would preferentially move if motion were triggered by a local disturbance. This information may therefore help interpret landing site stability and the mechanical context of future MMX surface observations.

\subsection{Assumptions and limitations}\label{Xsec17-6.3}

\label{Assumptions-Limitations}

This study is intentionally simplified, and several processes that could affect the Regolith Migration Pathways are not considered here. First, we do not include the effects of past orbital evolution. It is well established that the tidal evolution of Phobos is relatively rapid and that the satellite may have been located at significantly larger orbital distances in the past \citep{Jacobson_Lainey_2014} . Because Phobos is assumed to remain tidally locked, both tidal accelerations and centrifugal forces would have been weaker earlier in its history, potentially modifying the large-scale organization of the pathways. However, the scope of this work is deliberately restricted to the present-day configuration, using the current shape and topography rather than attempting to reconstruct past states. In this context, the low-threshold approach is intended to explore a relatively late stage of surface evolution, when the global shape of Phobos was already close to its present form, while local slopes and/or transient external perturbations could still permit enhanced transport. Attempting to reconstruct the past topography would be highly speculative at this stage, because neither the internal structure of Phobos nor the mechanical properties of the surface material are sufficiently constrained. The forthcoming MMX mission is expected to provide invaluable new data on both aspects.

We also do not include the effect of the forced libration of Phobos, whose amplitude is about 1.14$^\circ$ \citep{Rambaux_2012}{ . In principle, libration introduces an additional periodic forcing that could modulate surface accelerations. However, given its small amplitude compared to the dominant tidal and centrifugal effects, we expect its first-order influence to be limited. A dedicated assessment of this contribution is left for future work.

{ Because the present model is deliberately restricted to a kinematic framework, no assumptions are made about the internal structure of the surficial material, its rheological behaviour, or its detailed mechanical properties. In particular, the model does not resolve quantities or processes such as mass, cohesion, adhesion, inter-grain friction, rolling, collisions, or other grain-scale interactions. The trajectory calculations are instead formulated only in terms of the local acceleration field, the surface constraint, and a mobility-threshold criterion governing the initiation and cessation of motion. In addition, the local topography is assumed to be continuous and sufficiently resolved by the adopted digital terrain model, and the present-day acceleration field is used as the reference forcing framework during each trajectory integration. In that sense, the purpose of the model is not to reproduce the detailed physical behaviour of regolith, but to reveal the large-scale organization of Regolith Migration Pathways across the surface of Phobos. Future work may extend this framework toward more complete treatments of granular dynamics and contact mechanics.}

To reduce computational cost, trajectories are treated as surface-constrained (\linkref[Section]{\ref{section_surface_trajectories_and_different_friction_coefs}}). To assess the validity of this approximation, we also performed complementary simulations allowing temporary detachment from the surface and short ballistic excursions (\linkref[Appendix]{\ref{appendix:flying_particles_dynamics}}). A comparison between the surface-constrained trajectories (\linkref[\del{Figures}\ins{Figs.}]{\ref{fig:trajectories_14_deg_4panels}} and \ref{fig:trajectories_14_deg_poles}) and the detachment-allowed simulations (\linkref[\del{Figures}\ins{Figs.}]{\ref{fig:appendix_trajectories_flying_part_14_deg_4panels}} and \ref{fig:appendix_trajectories_flying_part_14_deg_north_south}) shows very similar large-scale Regolith Migration Pathways, with differences limited to localized depressions where trajectories may briefly leave the surface and re-contact nearby. Within the limits of our model, surface-constrained simulations therefore capture the first-order organization of Regolith Migration Pathways, while ballistic excursions mainly affect local-scale trajectory behaviour.

{ Another simplification is the assumption of a homogeneous interior for Phobos, which remains compatible with current radio-science constraints }\citep{Le_Maistre_2013, Le_Maistre_2019}{ , although the internal structure of Phobos is still an open question. By modifying the internal distribution of mascons inside Phobos} (\linkref[Appendix]{\ref{Appendix: acceleration}}){ , we verified that the main large-scale dynamical features and pathway patterns remain robust with respect to the variability introduced by the mascon discretization itself. This provides some indication that the global organization of the results is not dominated by numerical noise, although it does not replace a proper sensitivity study to realistic internal structure variations. We also note that Le Maistre {\rm et al.} (2019) concluded that a homogeneous and porous Phobos remains the simplest assumption compatible with the available constraints on libration amplitude and moments of inertia.}

{ Despite these simplifications, the methodology developed here provides a useful basis for future studies of surface transport on Phobos and other small bodies. Extending the present framework toward more realistic friction laws or more complete granular-dynamics treatments would help further constrain mobility under low-gravity conditions. Upcoming observations from the }\textit{MMX}{  mission will provide an opportunity to test these predictions against high-resolution topography, spectral data, and in situ grain analyses. Overall, this work establishes a methodological framework for interpreting surface transport on Phobos and for placing future MMX samples within their broader geomorphological and transport context.}

\PrintCredit

\begin{ack}
\section*{Acknowledgement}
The authors thank the two anonymous reviewers for their insightful comments, which significantly improved the quality of this manuscript. This research is supported by the French ANR project Roche, number ANR-23-CE49-0012, and French Space Agency (CNES). We thank Richard Binzel and Erik Asphaug for their useful comments and discussions.
\end{ack}

\begin{appgroup}

\appsection{Acceleration at the surface of Phobos} \label{Appendix: acceleration}

\def\thetable{A.\arabic{table}}
\setcounter{table}{0}
\def\thefigure{A.\arabic{figure}}
\setcounter{figure}{0}

{ In order to identify the Regolith Migration Pathways across the present-day topography of Phobos, we adopt a kinematic approach that accounts for the main acceleration components in the moon's rotating frame. The model computes the net acceleration vector projected onto the local surface plane at each point, thereby enforcing tangential motion constrained by the topography. This section outlines the contributing accelerations and the projection strategy.}

The total body acceleration \( \vec{a}_{\text{net}} \) acting on the surface includes the self-gravity of Phobos, the centrifugal acceleration linked to Phobos' rotation in the inertial frame, and the tidal acceleration due to Mars. { Once a trajectory is initiated}, the Coriolis acceleration must also be added.

Phobos' body frame is defined from the digital terrain model used here \citep{Ernst_2023}, with O located at the center of figure of Phobos (the origin of the digital terrain models of Phobos); the X-axis (\(\overrightarrow{e_x}\)) points toward Mars, the Z-axis (\(\overrightarrow{e_z}\)) points along the spin vector of Phobos (also perpendicular to Phobos' orbital plane), and the Y-axis (\(\overrightarrow{e_y} = \overrightarrow{e_z} \times \overrightarrow{e_x}\)) points approximately opposite to the direction of Phobos' motion (though not exactly, due to the eccentricity of Phobos' orbit).

As we work in a frame centered on the centroid of Phobos' DTM, where Phobos is fixed (a body-fixed frame), this frame is non-inertial. Therefore, the net { acceleration acting at a surface location} \(\vec{r}\) (measured with respect to the center of Phobos, O) and { with velocity} \(\vec{v}\) (in the body-fixed frame), can be written as follows:

\begin{equation}
 \vec{a_{net}}(\vec{r})=\vec{g}(\vec{r})+\vec{a_{c}}(\vec{r})+\vec{a_{t}}(\vec{r})+\vec{a_{cor}}(\vec{v})-\vec{a_{P-M}}(\vec{O})
\label{Xeqn1-A.1}
\end{equation}

They are computed as follows :

\begin{itemize}
    \item \( \vec{g} \): The local self-gravitational acceleration at each surface mesh node is obtained by summing the contributions from 10$^{6}$ equal-mass particles randomly and uniformly distributed within Phobos' volume, with a total mass equivalent to that of Phobos (this approach is sometimes referred to as a \textit{mascon method}). We have chosen this method because of its versatility, as it allows an easy modification of the internal mass distribution, which is poorly constrained for small bodies. However, in the case of a perfectly homogeneous interior,  it is less accurate  than the polyhedral method for a given surface mesh \citep{Scheeres_2010} . The accuracy of our approach is discussed in the next section.

    \item $\vec{a}_c=-\vec{\omega} \times ( \vec{\omega} \times \vec{r})$ is the centrifugal acceleration due to Phobos' rotation about its axis, where $\vec{\omega}$ is the spin vector of Phobos. In the body frame, where Phobos is fixed, by convention the Z axis is defined along Phobos' spin axis, so that $\vec{\omega}=\omega \overrightarrow{e_z}$ and where $\omega=0.0002279688$ rad/s is Phobos' rotational angular velocity, taken from the orbital period of Phobos (\citealp{Rambaux_2012}). We do not consider here the libration of Phobos that is about $1\degree$ in amplitude \citep{Rambaux_2012}.

    \item \( \vec{a}_t \) is the tidal term given by the gravitational pull from Mars at point $\vec{r}$ at the surface of Phobos. Let O be the center of Phobos, M the center of Mars and $\vec{A}=\overrightarrow{MO}+\vec{r}$ then
    $\vec{a}_t=-GM_{Mars}\vec{A}/\lVert  \vec{A} \rVert^3$.

    \item \( \vec{a}_{\text{cor}} = -2 \vec{\omega} \times \vec{v} \) is the Coriolis acceleration, where $\vec{\omega}$ is the spin vector of Phobos. { Because it depends on the instantaneous velocity, this term is included only during the dynamical integration of the surface trajectories}

    \item $\vec{a_{P-M}}=-GM_{Mars}\overrightarrow{MO}/\lVert  \overrightarrow{MO} \rVert^3$ is the instantaneous gravitational acceleration of Phobos' center (i.e. the origin of the reference frame) due to Mars. This term is an inertial pseudo-acceleration in Phobos' body-fixed frame. To compute the instantaneous location of Mars in Phobos' body frame, we assume that Phobos' orbital semi-major axis, eccentricity and inclination are a = 9375 km, e = 0.01511 \citep{Jacobson_Lainey_2014} and i = 0 $\deg$, respectively. The Kepler equation is solved to determine the instantaneous position of the center of Phobos, in the Mars-centered inertial frame, and then, converted to the Phobos rotating frame (with angular velocity $\omega$).

\end{itemize}

\begin{figure}

\centerline{\includegraphics[width=0.65\linewidth]{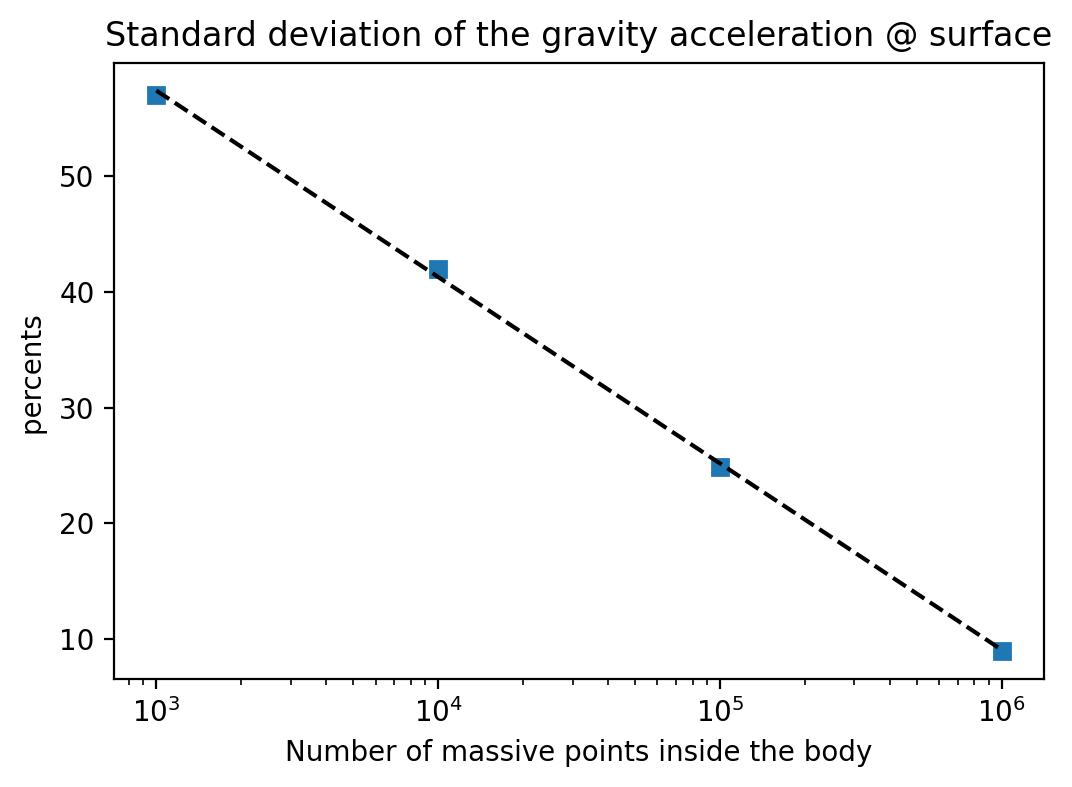}}
\caption{Variability in the computed norm of the surface gravity as a function of the number of mass points placed inside the body.}
     \label{fig:gravity_accuracy}
 \end{figure}

{ All acceleration terms that derive from a potential are computed at each node of a high-resolution ($\sim$36\del{ }\ins{\thinspace }m) triangular surface mesh} \citep{Ernst_2023} { and projected onto the local tangent plane, defined at each node as perpendicular to the local topographic normal } \( \vec{n} \). { Note that the Coriolis acceleration, which does not derive from a potential because it depends on velocity, is not included in these maps. The calculations are performed in a body-fixed frame centered on the centroid of Phobos' Digital Terrain Model} \citep{Ernst_2023}.

This approach follows standard practices in rotating-body dynamics while being adapted to the specific gravitational environment of Phobos.

\subsection{Projection onto the local tangent plane}\label{Xsec18-A.0.1}

In order to clearly visualize the direction of the local acceleration with respect to the local topography at every point on the surface, we introduce the local tangential acceleration \( \vec{a}_{\parallel} \), defined as follows. Let \( \vec{n} \) be the local topographic normal; then the local tangential acceleration is:
\begin{equation}
   \vec{a}_{\parallel} = \vec{a}_{\text{net}} - (\vec{a}_{\text{net}} \cdot \vec{n})\vec{n}
   \label{equation:a_surface}
\end{equation}

The study of \( \vec{a}_{\parallel} \) will provide, in addition to the dynamical slope (see~\linkref[Appendix]{\ref{subsec:slope_stability}}), a clear indication of the local direction in which regolith may be accelerated. Nevertheless, since the Coriolis force is not included at this stage, this still provides only a crude approximation of the local regolith motion, although much richer than the simple knowledge of the local dynamical slope.

We will discuss the implications and interpretation of the direction of \( \vec{a}_{\parallel} \) in the next section.

\appsection{Computation of surface trajectories} \label{Appendix:trajectories}

\def\thetable{B.\arabic{table}}
\setcounter{table}{0}
\def\thefigure{B.\arabic{figure}}
\setcounter{figure}{0}

\subsection{Physical assumptions}\label{Xsec19-B.1}

In RAVEL, the surface mobility of regolith on Phobos is treated as a deterministic response to the effective acceleration field acting on each surface element. The effective acceleration, $\mathbf{a}_{\text{eff}}$, combines the gravitational, centrifugal and tidal components, which are precomputed for a fixed orbital configuration and projected onto the local tangent plane of the topography. Hence, the dynamical analysis is performed in the body-fixed frame of Phobos, assuming the moon remains static at a given position in its orbit.

Each surface element is characterized by its local topographic normal vector $\mathbf{n}$ and the tangent plane defined by it. The total acceleration acting on a surface element can be decomposed into tangential and normal components: the tangential part contributes to the motion along the surface, whereas the normal part modifies the apparent normal load and therefore the frictional resistance. The motion at the surface is thus governed by the tangential component of the effective acceleration and by the Coriolis term, which depends on the local instantaneous velocity and Phobos' rotation rate $\boldsymbol{\omega}$.

\subsection{Governing equations}\label{Xsec20-B.2}

We describe the motion at the surface of Phobos using a first-order kinematic system of equations, given by
\begin{equation}
\frac{d\mathbf{r}}{dt} = \mathbf{v},
\label{Xeqn3-B.1}
\end{equation}
\begin{equation}
\frac{d\mathbf{v}}{dt} = \mathbf{a}_{\text{T}},
\label{Xeqn4-B.2}
\end{equation}
where $\mathbf{r}$ is the position vector and $\mathbf{v}$ is the local instantaneous velocity at position $\mathbf{r}$. The total local acceleration acting on the surface is expressed as
\begin{equation}
\mathbf{a}_{\text{T}} = \mathbf{a}_{\text{eff}} + \mathbf{a}_{\text{cor}} + \mathbf{a}_{\text{fric}},
\label{eq:eom}
\end{equation}
where $\mathbf{a}_{\text{eff}}$ is the effective acceleration (gravitational + centrifugal + tidal), $\mathbf{a}_{\text{cor}}$ is the Coriolis term, and $\mathbf{a}_{\text{fric}}$ is the dynamic frictional deceleration.

The Coriolis acceleration is given by
\begin{equation}
\mathbf{a}_{\text{cor}} = -2 \, \boldsymbol{\omega} \times \mathbf{v},
\label{eq:a_cor}
\end{equation}
where $\boldsymbol{\omega}$ is Phobos' spin rate. Thus, the normal component $(\mathbf{a}_{\text{cor}}\cdot \mathbf{n})$ modifies the frictional load. The dynamic friction law follows a Coulomb-type relationship, in which the frictional acceleration opposes the direction of motion and is proportional to the effective normal load:
\begin{equation}
\mathbf{a}_{\text{fric}} = - \tan(\phi_{\text{dynamic}})\, \left( a_{\text{eff},n} + a_{\text{cor},n} \right) \frac{\mathbf{v}_{\text{proj}}}{|\mathbf{v}_{\text{proj}}|},
\label{Xeqn7-B.5}
\end{equation}
where $a_{\text{eff},n} = \mathbf{a}_{\text{eff}}\cdot \mathbf{n}$ and $a_{\text{cor},n} = \mathbf{a}_{\text{cor}}\cdot \mathbf{n}$ are the normal components of the effective and Coriolis local accelerations, respectively, and
\[
\mathbf{v}_{\text{proj}} = \mathbf{v} - (\mathbf{v}\cdot \mathbf{n})\,\mathbf{n}
\]
is the velocity vector projected onto the local tangent plane.

The dynamic friction angle, $\phi_{\text{dynamic}}$, represents the kinetic regime and is assumed to be smaller than the static friction angle, $\phi_{\text{static}}$, of the material.

The motion at the surface is driven by the projection of the total local acceleration onto the local tangent plane of the topography:
\[
\mathbf{a}_{\text{T,proj}} = \mathbf{a}_{\text{T}} - (\mathbf{a}_{\text{T}}\cdot \mathbf{n})\,\mathbf{n}.
\]

\subsection{Local dynamical slope and friction criteria}\label{Xsec21-B.3} \label{dynamical-slope}

Let $\mathbf{a}_{\text{eff}}$ be the effective local acceleration (including gravitational, centrifugal and tidal components), and let $\mathbf{n}$ denote the outward unit normal to the local surface. The angle $\theta$ between $\mathbf{a}_{\text{eff}}$ and $\mathbf{n}$ is defined through
\[
\cos \theta = \frac{\mathbf{a}_{\text{eff}} \cdot \mathbf{(-n)}}{\|\mathbf{a}_{\text{eff}}\| \,\|\mathbf{n}\|}.
\]
{ Thus, the local \textit{dynamical slope} can be written as}
\begin{equation} \label{dyn_slope}
\theta = \pi - \arccos\left( \frac{\mathbf{a}_{\text{eff}} \cdot \mathbf{n}}{\|\mathbf{a}_{\text{eff}}\|} \right).
\end{equation}

We distinguish between a \textit{static} and a \textit{dynamic friction angle}. Trajectories can only start from points where the dynamical slope exceeds the \textit{static friction angle}
\[
\theta > \phi_{\text{static}},
\]
and propagation continues only through regions where
\[
\theta > \phi_{\text{dynamic}}.
\]

{ This distinction is consistent with the general behavior of granular materials, for which dynamic friction is commonly lower than static friction.

Phobos' surface may be dominated by a fine-grained, weakly consolidated regolith }\citep{fornasier2024, Smith2018, Miyamoto_2021}{ . In such low-cohesion granular media, the angle of repose may closely approximate the static friction angle $\phi_{\text{static}}$ }\citep{Cardena-Barrantes2025}{ . Surface slope distributions on Eros, Itokawa, Ryugu, and Bennu rarely exceed 30$^\circ$, and tend to cluster below the estimated static friction angle }\citep{Barnouin2019}{ . It was found that Bennu's current shape can be maintained with an internal friction angle as low as $\sim 18 ^{\circ}$ , close to its average surface slope of $\sim 17 ^{\circ}$ }\citep{Barnouin2019}{ . Likewise, lunar surface and simulation studies report repose angles of 25$^\circ$ to 35$^\circ$, which correspond to in-situ and laboratory estimates} \citep{Carrier1991}{ . However, in the present study we do not attempt to infer the intrinsic frictional properties of Phobos' regolith from such analogies.

In the case of Phobos, the observed mean dynamical surface slope is about $14^\circ$ (see \linkref[Fig.]{\ref{fig:slope_histrogram}})}{ . In our model, this value is not interpreted as a direct estimate of the intrinsic inter-grain friction of the regolith. Rather, it is used as a low-threshold reference level within the present-day topography, allowing us to reveal the broader organization of \textit{Regolith Migration Pathways} when past topography and material properties remain unconstrained. In that sense, the $14^\circ$ case is intended as a pathway-revealing end-member, representative of conditions of enhanced mobility, whether associated with steeper local slopes in the past or with transport triggered by external perturbations such as impact shaking or tidal stressing.}

\subsection{Initial conditions}\label{Xsec22-B.4}

The integration begins from a discrete set of starting points distributed across the surface. At these locations, the material is initially at rest,
\[
\mathbf{v}(t=0) = 0,
\]
and motion can only start at those points where the friction criterion $\theta > \phi_{\text{static}}$ is satisfied. Once the motion is triggered, the tangential acceleration is updated iteratively, and the trajectory is advanced only through regions where $\theta > \phi_{\text{dynamic}}$, until the termination criterion defined in \linkref[Section]{\ref{termination-criterion}} is met.

\subsection{Numerical integration scheme}\label{Xsec23-B.5}

The motion is integrated using an adaptive explicit Euler scheme that ensures numerical stability. The time step $\Delta t$ is defined as
\[
\Delta t = \min(\Delta t_0, \, C_{\text{safe}} \, \frac{\Delta R}{|\mathbf{v}_{\text{proj}}|}),
\]
where $\Delta t_0$ is the initial time step, $\Delta R$ is the mean spatial resolution of the surface mesh (typically corresponding to the resolution of the DTM), $\mathbf{v}_{\text{proj}}$ is the velocity projected on the tangent plane, and $C_{\text{safe}}$ is a safety factor typically below unity.

At each step, the velocity and position are updated using an explicit Euler integration scheme:
\begin{align}
\mathbf{v}^{t+1}_{\text{proj}} &= \mathbf{v}^{t}_{\text{proj}} + \mathbf{a}_{\text{T,proj}}  \, \Delta t, \\
\mathbf{r}^{t+1}_{\text{proj}} &= \mathbf{r}^{t}_{\text{proj}} + \mathbf{v}^{t+1}_{\text{proj}} \, \Delta t,
\end{align}
where $\mathbf{a}_{\text{T,proj}}$ is the total local acceleration projected onto the local tangent plane and $\mathbf{v}^{t+1}_{\text{proj}} = \mathbf{v}^{t+1} - (\mathbf{v}^{t+1}\cdot \mathbf{n})\,\mathbf{n}$ enforces the surface constraint.

At each integration step, the motion is constrained to remain within the local tangent plane to the surface: the displacement vector $\mathrm{d}\mathbf{r}$ is enforced to be orthogonal to the surface normal $\mathbf{n}$ at all times. In other words, the motion is constrained to remain bound to the surface and detachment is not allowed. The sequence of updated positions thus defines surface-constrained trajectories. The validity of this approximation, as well as comparisons with the trajectories that are free to detach from the surface, are discussed in \linkref[Section]{\ref{Assumptions-Limitations}} and in \linkref[Appendix]{\ref{appendix:flying_particles_dynamics}}.

\subsection{Energy-based termination criterion}\label{Xsec24-B.6}
\label{termination-criterion}

To decide when a trajectory has effectively come to rest, we use an energy-based stopping criterion applied to the tangential motion. We define the specific tangential kinetic energy at step $t$ as
\[
e_{\text{k}}^{\,t} = \tfrac{1}{2}\,\big\|\mathbf{v}^{\,t}_{\text{proj}}\big\|^{2},
\]

We define the projected displacement between two consecutive steps as
\[
\Delta \mathbf{r}^{\,t}_{\text{proj}} = \mathbf{r}^{\,t+1}_{\text{proj}} - \mathbf{r}^{\,t}_{\text{proj}}.
\]
A discrete work-energy balance for the tangential motion then gives the predicted specific kinetic energy at step $t+1$:
\begin{equation}
e_{\text{k}}^{\,t+1,\text{pred}}
=
e_{\text{k}}^{\,t}
+ \mathbf{a}^{\,t}_{\text{T,proj}} \cdot \Delta \mathbf{r}^{\,t}_{\text{proj}}.
\label{Xeqn9-B.9}
\end{equation}

Once a finite tangential velocity has been established, if
\begin{equation}
\label{eq:ek_stop}
e_{\text{k}}^{\,t+1,\text{pred}} < 0,
\end{equation}
the discrete work-energy relation would imply a negative specific kinetic energy, which is unphysical. We therefore interpret this as indicating that local equilibrium (i.e. vanishing tangential kinetic energy) has been reached before completion of the step, and terminate the trajectory at the last valid position $\mathbf{r}^{\,t}_{\text{proj}}$.

\appsection{Maps of slope stability and eccentricity effect}
\label{subsec:slope_stability}

\def\thetable{C.\arabic{table}}
\setcounter{table}{0}
\def\thefigure{C.\arabic{figure}}
\setcounter{figure}{0}

\subsection{A-priori mobility map from orbital dynamics}\label{Xsec25-C.1}

To characterize surface stability and regolith mobility on Phobos, we first derive an a-priori mobility map from the underlying orbital dynamics. The instantaneous acceleration field across the surface along Phobos' eccentric orbit around Mars, is converted into dynamic slopes and then associated to a mobility component.
 These accelerations combine the body's self-gravity, centrifugal acceleration from its spin, and the time-varying tidal acceleration from Mars, evaluated on the irregular topography and projected into the local surface frame. For each surface element $i$ and for each orbital configuration considered here, pericenter ($M=0^\circ$) and apocenter ($M=180^\circ$), we compute the \emph{dynamic slope} $\theta$ as
\begin{equation}
\tan\theta_i^{(k)} =
\frac{\left\lVert \mathbf{a}^{(k)}_{i,\mathrm{tan}} \right\rVert}
     {a^{(k)}_{i,\mathrm{norm}}}\,,
\label{Xeqn11-C.1}
\end{equation}
where $\mathbf{a}^{(k)}_{i,\mathrm{tan}}$ is the tangential component of the net acceleration and $a^{(k)}_{i,\mathrm{norm}}$ its normal component for orbital state $k \in \{0^\circ, 180^\circ\}$. This yields two global slope fields, $\theta_i^{(0)}$ and $\theta_i^{(180)}$.

From these two fields we derive three scalar indicators that quantify, for each surface element, (i) how steep it is during the orbit, (ii) how close it lies to a long-term equilibrium configuration, and (iii) how strongly it is modulated by orbital forcing. These indicators are then transformed into dimensionless \emph{mobility components} and combined into a single a-priori \emph{mobility index} $M_i \in [0,1]$ (\linkref[Fig.]{\ref{fig:mobility_map}}). As discussed below in the trajectory-based analysis, regions with large values of these components systematically correspond to long, high-mobility  trajectories.

\paragraph*{Maximum dynamic slope}
The first indicator is the maximum dynamic slope
\begin{equation}
\theta_{i,\max} = \max\!\left(\theta_i^{(0)}, \theta_i^{(180)}\right),
\label{Xeqn12-C.2}
\end{equation}
which measures whether a given location becomes transiently very steep at any point of the orbit.

\paragraph*{Mean dynamic slope and global reference level}
For each facet we define its local mean,
\begin{equation}
\bar{\theta}_i = \frac{1}{2}\left(\theta_i^{(0)} + \theta_i^{(180)}\right),
\label{Xeqn13-C.3}
\end{equation}
and we also compute a \emph{global orbital-mean slope}
\begin{equation}
\theta_{\mathrm{mean}} = \left\langle \bar{\theta}_i \right\rangle_i,
\label{Xeqn14-C.4}
\end{equation}
which acts as a proxy for a stability condition of the surface under cyclic forcing. Regions with $\bar{\theta}_i \gtrsim \theta_{\mathrm{mean}}$ are, on average, closer to failure.

\paragraph*{Temporal variability of the dynamic slope}
The third indicator quantifies the amplitude of the orbital modulation at each location,
\begin{equation}
\Delta\theta_i = \left|\theta_i^{(180)} - \theta_i^{(0)}\right|,
\label{Xeqn15-C.5}
\end{equation}
and captures how strongly time-varying forcing perturbs local stability.

\subsubsection{Steep-slope mobility component}\label{Xsec26-C.1.1}

The first mobility component, $C^{\mathrm{steep}}_i$, measures how much the maximum slope at a given location exceeds the global mean. We distinguish between surfaces that are consistently steep throughout the orbit and those that are only steep during one orbital phase:
\begin{equation}
C^{\mathrm{steep}}_{i} =
\begin{cases}
\dfrac{2\,[\theta_{i,\max} - \theta_{\mathrm{mean}}]}{\theta_{\mathrm{mean}}},
& \theta_i^{(0)} > \theta_{\mathrm{mean}} \ \text{and} \
  \theta_i^{(180)} > \theta_{\mathrm{mean}}, \\[0.8em]
\dfrac{\theta_{i,\max} - \theta_{\mathrm{mean}}}{\theta_{\mathrm{mean}}},
& \text{exactly one of } \{\theta_i^{(0)}, \theta_i^{(180)}\}
  > \theta_{\mathrm{mean}}, \\[0.4em]
0, & \text{otherwise}.
\end{cases}
\label{Xeqn16-C.6}
\end{equation}
Thus, locations that are always steeper than the global mean receive twice the weight of those that only exceed it part of the time. Finally, $C^{\mathrm{steep}}_{i}$ is normalized by its maximum over the body so that $\max_i C^{\mathrm{steep}}_{i} = 1$.

\subsubsection{Cyclic-slope mobility component}\label{Xsec27-C.1.2}

The second mobility component, $C^{\mathrm{var}}_i$, captures the \emph{relative} orbital modulation of the slope,
\begin{equation}
C^{\mathrm{var}}_{i} =
\frac{\Delta\theta_i}{\theta_{\mathrm{mean}}},
\label{Xeqn17-C.7}
\end{equation}
again rescaled by its maximum so that $\max_i C^{\mathrm{var}}_{i} = 1$. Large values identify locations where the effective driving component of gravity changes significantly over an orbital cycle, even if the mean slope is moderate. Such regions are expected to favour creep and diffusive mass redistribution. In practice, however, the a-posteriori trajectory analysis shows that the explicit effect of Phobos' orbital eccentricity on the transport pathways is generally negligible; this component should therefore be interpreted as a measure of generic cyclic forcing rather than of eccentricity alone.

\subsubsection{Critical-instability component}\label{Xsec28-C.1.3}

The third component, $C^{\mathrm{crit}}_i$, is an ordinal index that emphasizes \emph{synergistic combinations} of high average slope and strong orbital modulation. We first define a variation threshold
\begin{equation}
\Delta\theta_{\mathrm{thr}} = 0.5 \,\max_i (\Delta\theta_i),
\label{Xeqn18-C.8}
\end{equation}
and then classify each facet into one of several mutually exclusive categories according to whether its slopes are above or below $\theta_{\mathrm{mean}}$ at the two orbital phases, and whether $\Delta\theta_i$ exceeds $\Delta\theta_{\mathrm{thr}}$. Numerically, these categories are encoded as integer levels from 0 to 5 (from least to most unstable):
\begin{itemize}
  \item level 5: both $\theta_i^{(0)}$ and $\theta_i^{(180)}$ exceed $\theta_{\mathrm{mean}}$ and $\Delta\theta_i > \Delta\theta_{\mathrm{thr}}$,
  \item level 4: one slope above $\theta_{\mathrm{mean}}$ and $\Delta\theta_i > \Delta\theta_{\mathrm{thr}}$,
  \item levels 3--1: progressively less demanding combinations with high variability but lower mean slopes,
  \item level 0: both slopes below $\theta_{\mathrm{mean}}$ and low variability.
\end{itemize}
The resulting discrete field $C^{\mathrm{crit}}_{i}$ is finally normalized to $[0,1]$ by division by its maximum. High values of $C^{\mathrm{crit}}_{i}$ highlight areas where steepness and temporal modulation jointly push the surface close to failure conditions.

\subsubsection{Combined mobility index}\label{Xsec29-C.1.4}

The \emph{mobility index} $M_i$ used in the map of \linkref[Fig.]{\ref{fig:mobility_map}} is obtained as a weighted linear combination of the three normalized mobility components,
\begin{equation}
M_i =
      w_1 \,C^{\mathrm{steep}}_{i} +
      w_2 \,C^{\mathrm{var}}_{i} +
      w_3 \,C^{\mathrm{crit}}_{i},
\label{Xeqn19-C.9}
\end{equation}
with equal weights $w_1 = w_2 = w_3 = 0.33$. Because each term is individually normalized, $M_i$ lies in the interval $[0,1]$ up to numerical rounding. Values of $M_i$ close to 1 mark terrain that is persistently steep and subject to strong cyclic forcing, and are therefore most susceptible to regolith motion; values near 0 correspond to gently sloping, weakly varying regions that are expected to act as long-term accumulation areas. In the trajectory-based analysis presented below, these high-$M_i$ regions are indeed found to coincide with starting points that generate the longest and most mobile surface trajectories.

\begin{figure*}

\centerline{\includegraphics[width=0.99\textwidth]{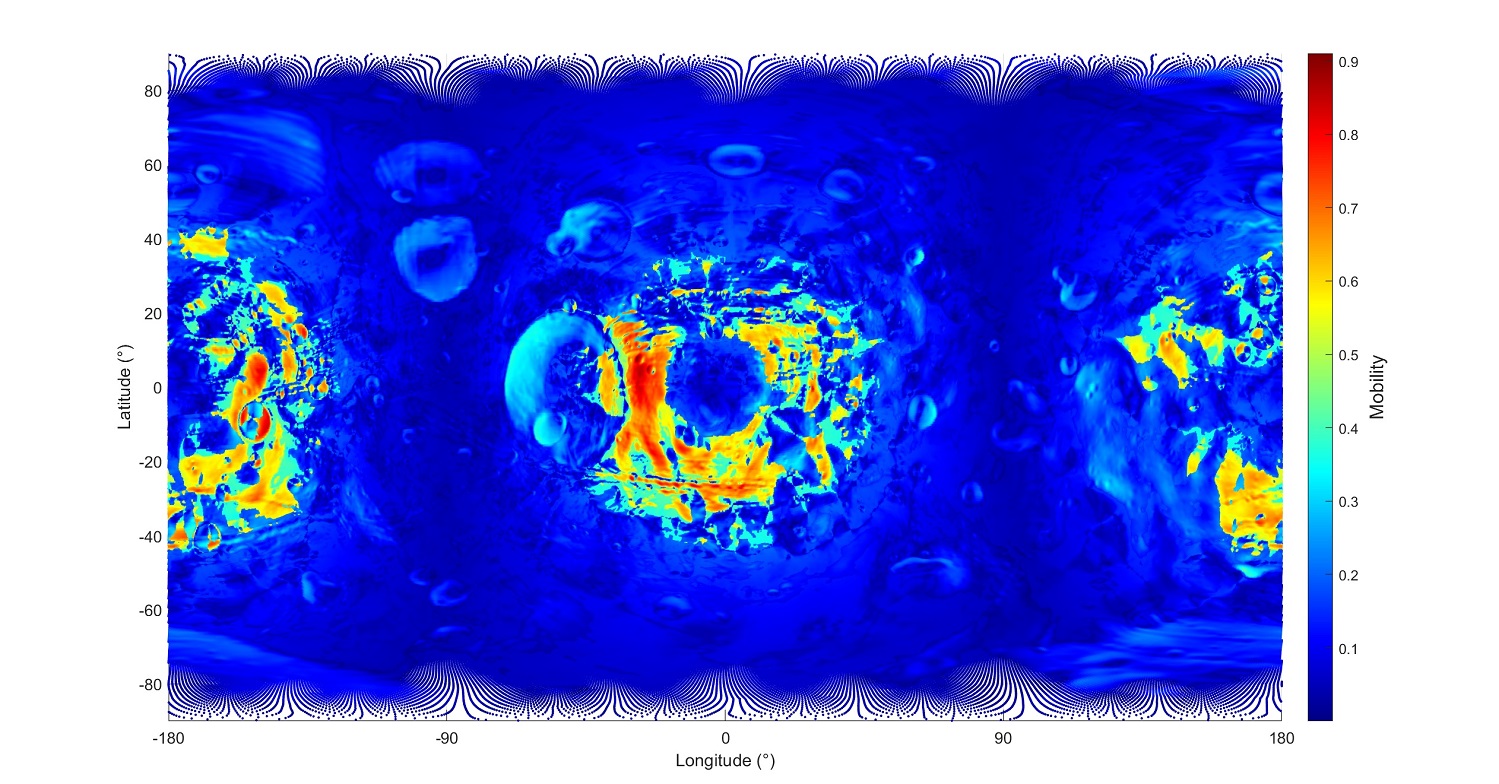}}
\caption{A-priori surface mobility map on Phobos. Colours show the normalized mobility index $M_i$ (warm colours for high mobility, cool colours for low mobility), derived from the combination of maximum dynamic slope, mean dynamic slope, and temporal variability between pericenter and apocenter.}
  \label{fig:mobility_map}
\end{figure*}

\subsection{A-posteriori mobility from trajectories computation}\label{Xsec30-C.2}

To complement this a-priori stability assessment, we explicitly integrate the surface-constrained trajectories, starting from $N_{\mathrm{start}} = 5000$ initial locations distributed all around the surface. For each starting point we compute two trajectories, corresponding to pericenter and apocenter conditions, using the same acceleration fields described above, a Coulomb-type friction law with same static and dynamic friction angles ($=14^{\circ}$), and a surface-constrained Euler scheme with adaptive time step and Coriolis correction.

We introduce an \emph{eccentricity--effect index} $E_i$ designed to quantify, for each starting location, how strongly orbital eccentricity modifies the distribution of possible trajectory endpoints associated with that location:
\begin{enumerate}
  \item For each starting point $i$, we consider the endpoints of the pericenter and apocenter trajectories.
  \item If two distinct endpoints exist, we define their mutual separation in 3D,
  \begin{equation}
  d_{ee} = \left\lVert \mathbf{e}_1 - \mathbf{e}_2 \right\rVert.
  \label{Xeqn20-C.10}
\end{equation}
  \item Let $\mathbf{p}_0$ be the initial position. We compute the start--to--endpoint distances,
  \begin{equation}
  d_1 = \left\lVert \mathbf{e}_1 - \mathbf{p}_0 \right\rVert,\qquad
  d_2 = \left\lVert \mathbf{e}_2 - \mathbf{p}_0 \right\rVert,
  \label{Xeqn21-C.11}
\end{equation}
  and their mean
  \begin{equation}
  d_{se} = \frac{1}{2}\left(d_1 + d_2\right).
  \label{Xeqn22-C.12}
\end{equation}
  \item The eccentricity--effect index is then defined as the dimensionless ratio
  \begin{equation}
  E_i = \frac{d_{ee}}{d_{se}} \leq 2,
  \label{Xeqn23-C.13}
\end{equation}
 with an additional safeguard: starting points with $d_{se}$ below a small tolerance are treated as immobile and are omitted from this analysis.

\end{enumerate}

When evaluated with the paired pericenter/apocenter trajectories, the eccentricity--effect index exhibits a broad, strongly right-skewed distribution. The resulting index spans nearly four orders of magnitude, from $E_i \simeq 3.6\times10^{-4}$ up to $E_i \simeq 1.91$. The median value is $E_i \approx 5.7\times10^{-2}$, while the mean is higher ($\approx 1.27\times10^{-1}$), reflecting a long tail of comparatively large eccentricity effects.

A more robust summary is provided by the cumulative distribution function (CDF) shown in \linkref[Fig.]{\ref{CDF_ecc}}. By construction, quartiles fall at $E_i \approx 0.0227$ (25th percentile), $E_i \approx 0.0571$ (median), and $E_i \approx 0.114$ (75th percentile). Therefore, for 75\% of the valid starting points, the pericenter--apocenter endpoint separation remains below $\sim0.11$ times the typical excursion distance from the start to the endpoints. In contrast, a minority of locations populate the high-$E_i$ tail (up to $1.9$), indicating that for those starts the pericenter/apocenter endpoints can become separated by a distance comparable to, or larger than, the characteristic start--to--endpoint excursion.

The spatial distribution of $E_i$ over the surface is shown in \linkref[Fig.]{\ref{fig:ecc_map}}, where starting points are coloured by quartiles of the eccentricity effect index over the dynamical slope background. Low values (Q1--Q2) are widespread and dominate the high-latitude regions, consistent with trajectories that are comparatively insensitive to the pericenter/apocenter forcing change. High-$E_i$ locations (Q4: $E_i \gtrsim 0.114$) are not randomly distributed; they cluster along several dynamically active corridors and around major morphodynamic structures, suggesting that eccentricity primarily acts as a \emph{route-selection} modifier in regions where the surface dynamics already admits multiple competing transport paths. In such areas, modest changes in the acceleration field between pericenter and apocenter can shift material into different basins of attraction, producing substantially different deposition sites. A limiting case occurs when the local dynamical slope is close to the critical threshold: a configuration that remains stable at pericenter may become unstable at apocenter (or vice versa), leading to motion in only one orbital phase and therefore to the largest endpoint separations and $E_i$ values approaching 2.

Overall, these results indicate that Phobos' orbital eccentricity is \emph{not} dynamically irrelevant in the trajectory-based (a-posteriori) transport problem: while the dominant behaviour is a weak eccentricity imprint for most starting points, a non-negligible subset of the surface exhibits strong sensitivity, with markedly distinct pericenter and apocenter deposition outcomes. Therefore, eccentricity has a secondary but spatially localised control on regolith redistribution, which is superimposed on the primary controls exerted by the shape-induced gravity field, rotation, tides, and local topography.

\begin{figure*}

\centerline{\includegraphics[width=0.99\textwidth]{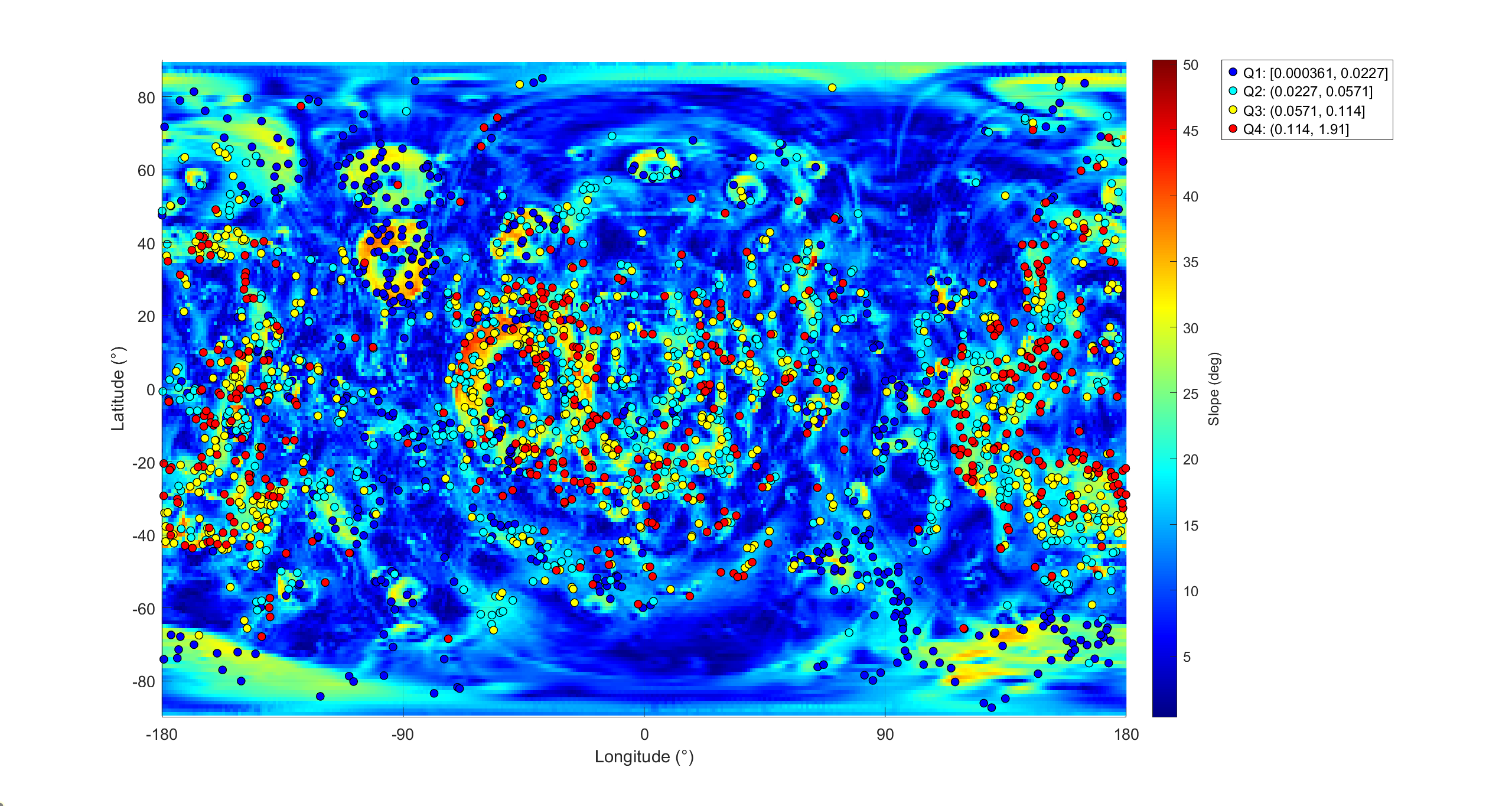}}
\caption{A-posteriori eccentricity index map. Starting points of the surface trajectories are plotted over the dynamical slope map and coloured by quartiles of the eccentricity index $E_i$, from blue (lowest values) to red (highest eccentricity effect). This map provides a direct measure of the influence of orbital eccentricity on the final depositional location.}
  \label{fig:ecc_map}
\end{figure*}

\begin{figure*}

\centerline{\includegraphics[width=0.95\linewidth]{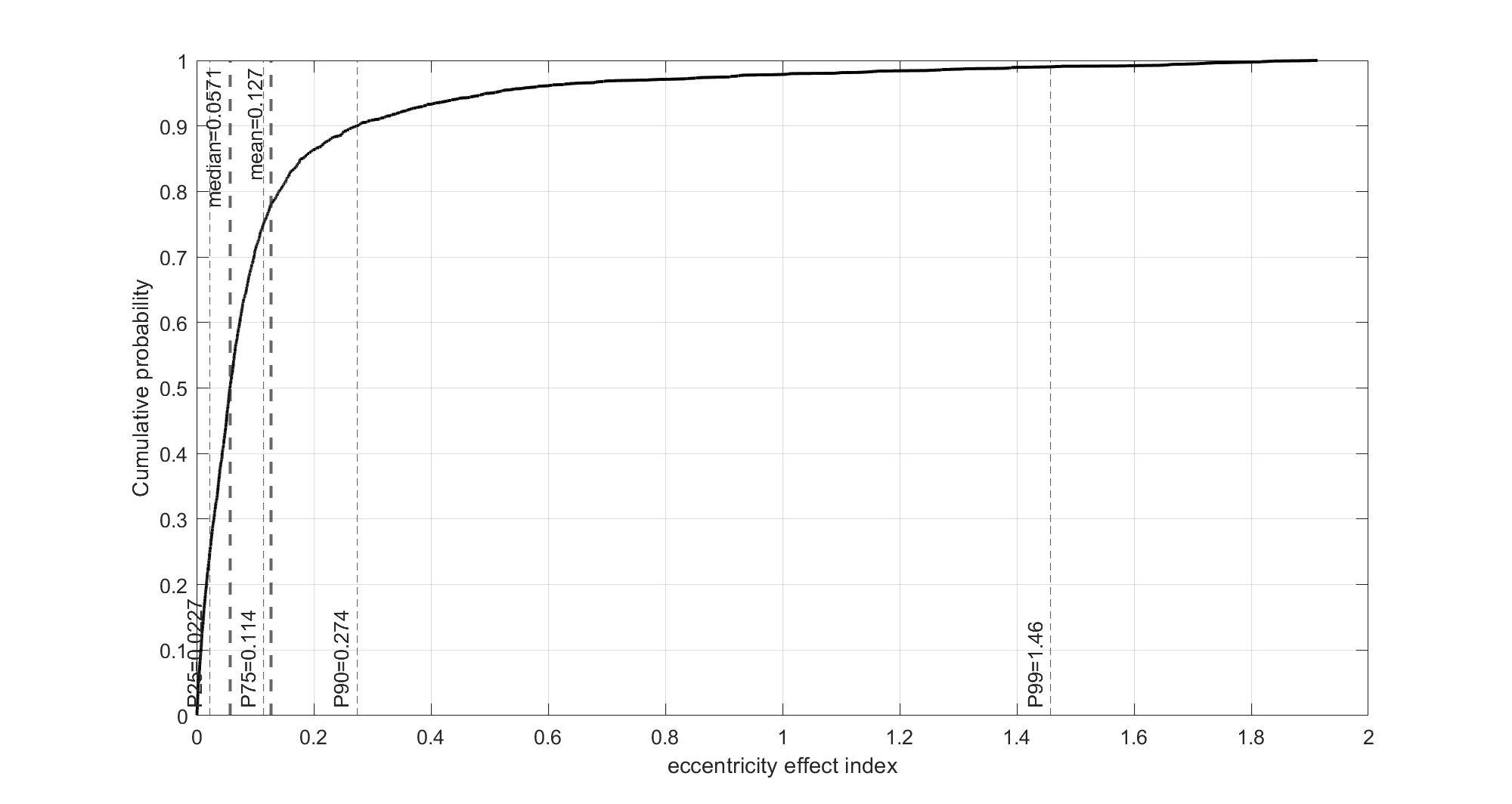}}
\caption{Cumulative distribution function (CDF) of eccentricity index $E_i$. The distribution is strongly right-skewed, with a median $E_i \simeq 5.7\times 10^{-2}$ and a long tail reaching $\sim 1.9$, indicating that a non-negligible subset of starting locations exhibits strong
sensitivity, with markedly distinct pericenter and apocenter deposition outcomes.}
  \label{CDF_ecc}
\end{figure*}

In addition to the eccentricity--effect index $E_i$, we quantify the effective mobility at each starting location in terms of the \emph{total path length} of the corresponding surface trajectory. For each simulated trajectory we compute the accumulated travelled distance along the surface, $D_i$ (in metres), which provides a direct a-posteriori measure of how far material can be transported under the combined action of effective gravity, Coriolis accelerations, and the adopted friction law. Since the eccentricity-induced divergence between pericenter and apocenter trajectories is small in most cases (\linkref[Fig.]{\ref{CDF_ecc}}), $D_i$ can be interpreted as a reliable indicator for mobility that is not sensitive to orbital phase. Note that, in our analysis, we exclude starting locations from which no trajectory is initiated, i.e. locations where the local dynamical slope remains below the low-threshold end-member ($14^{\circ}$ here). Only starting locations from which motion is initiated are considered in the discussion.

In the same spirit, it is useful to compare the duration of the simulated surface transport with the orbital timescale: Phobos completes one revolution around Mars in $\sim$7.65~h, whereas the computed trajectory durations are typically much shorter (\linkref[Fig.]{\ref{fig:trajtime_cdf}}). The CDF of total trajectory time shows a median duration of $\simeq 0.68$~h and a mean of $\simeq 0.81$~h, indicating that most trajectories that initiate motion come to rest within $\sim$10\% of an orbital period. Even the longest trajectories remain below $\sim$4~h (i.e., shorter than about half an orbit). Therefore, in the majority of cases, surface transport is effectively completed within a limited fraction of the orbital cycle, supporting the interpretation of $D_i$ as a reliable global mobility indicator. At the same time, the rare trajectories lasting several hours should be treated cautiously in local analyses, as eccentricity can induce a non-negligible evolution of the acceleration field over the duration of a single transport event.
\begin{figure*}

\centerline{\includegraphics[width=0.95\linewidth]{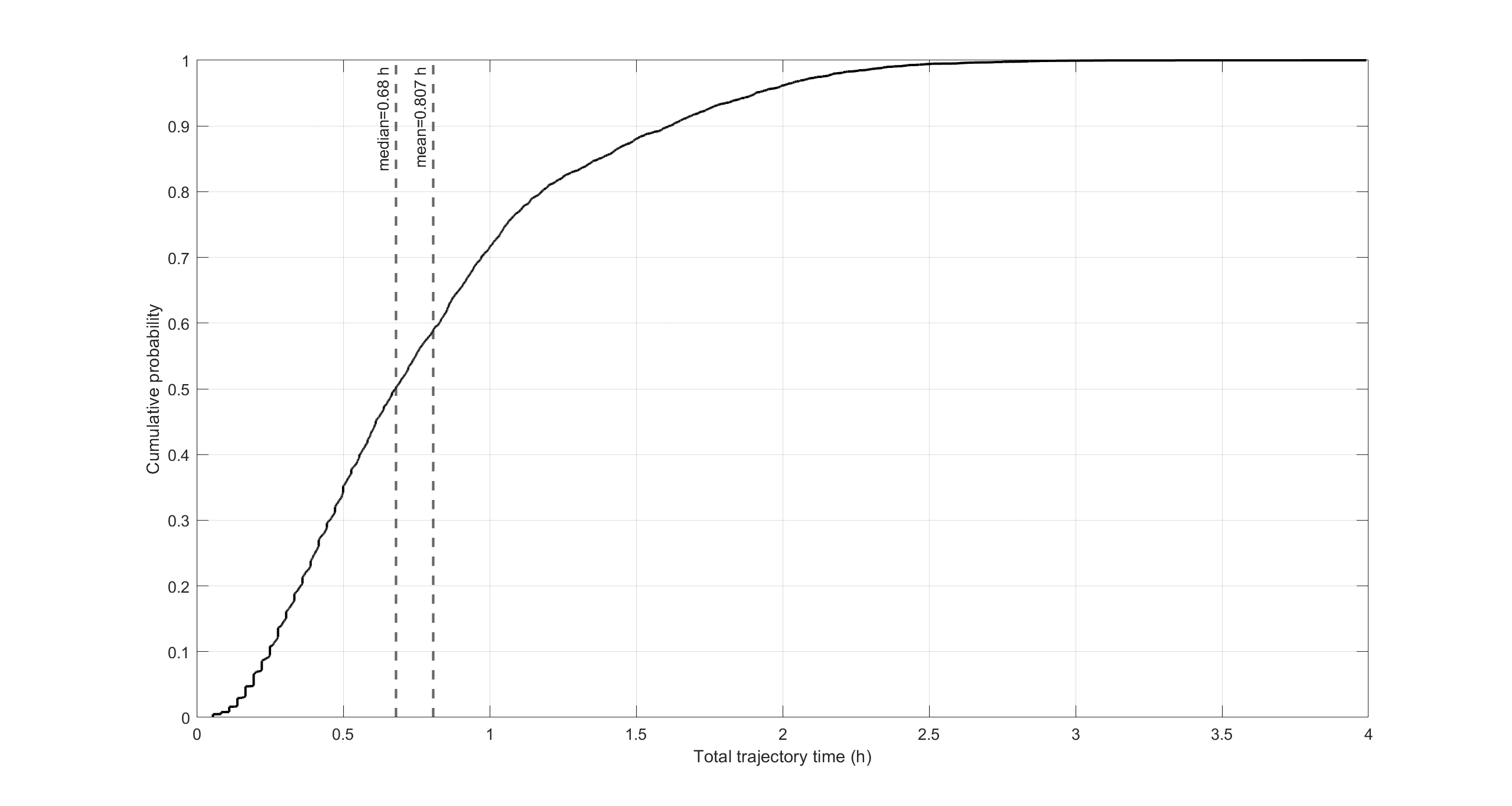}}
\caption{Cumulative distribution function (CDF) of total trajectory duration (in hours). The distribution shows a median duration of $\simeq 0.68$~h and a mean of $\simeq 0.81$~h, with the longest trajectories reaching $\sim$4~h. Compared with Phobos' orbital period around Mars ($\sim$7.65~h), most simulated surface-transport events are completed within a small fraction of an orbit.}
  \label{fig:trajtime_cdf}
\end{figure*}

To analyse the spatial distribution of trajectory lengths, we align the values of $D_i$ with the geographic coordinates of their starting locations and compute quartiles. The starting points are then plotted over the dynamical slope map and coloured according to the quartile to which their total path length belongs (\linkref[Fig.]{\ref{fig:trajlength_map}}).

Short-path trajectories (blue, $Q_1$) correspond to locally confined motions, consistent with immediate trapping in small-scale depressions or rapid frictional arrest. In contrast, long-path trajectories (red, $Q_4$) highlight starting locations from which trajectories can traverse kilometre-scale distances, exploring a large fraction of the surrounding terrain before coming to rest. Importantly, high-$D_i$ starting points are not uniformly distributed but cluster in distinct geographic bands and around specific morphologic structures, indicating that long-range transport is channeled by the interplay between topography and the global acceleration field rather than being a simple monotonic function of local slope.

The distribution of $D_i$ is strongly right-skewed (\linkref[Fig.]{\ref{fig:trajlength_cdf}}). The CDF shows a median trajectory length of $\sim 1.3\times 10^3$~m, implying that half of the (finite) trajectories travel more than $\sim$1~km along the surface. The curve rises rapidly at low values and then develops a long tail extending up to $\sim 1.4\times 10^4$~m, demonstrating that a relatively small fraction of starting locations produce exceptionally long transport paths. These outliers are spatially coherent in the quartile map (\linkref[Fig.]{\ref{fig:trajlength_map}}), supporting the interpretation that Phobos' surface contains preferential dynamical corridors where regolith can be redistributed over kilometre scales, while most regions remain comparatively confined.

\begin{figure*}

\centerline{\includegraphics[width=0.95\linewidth]{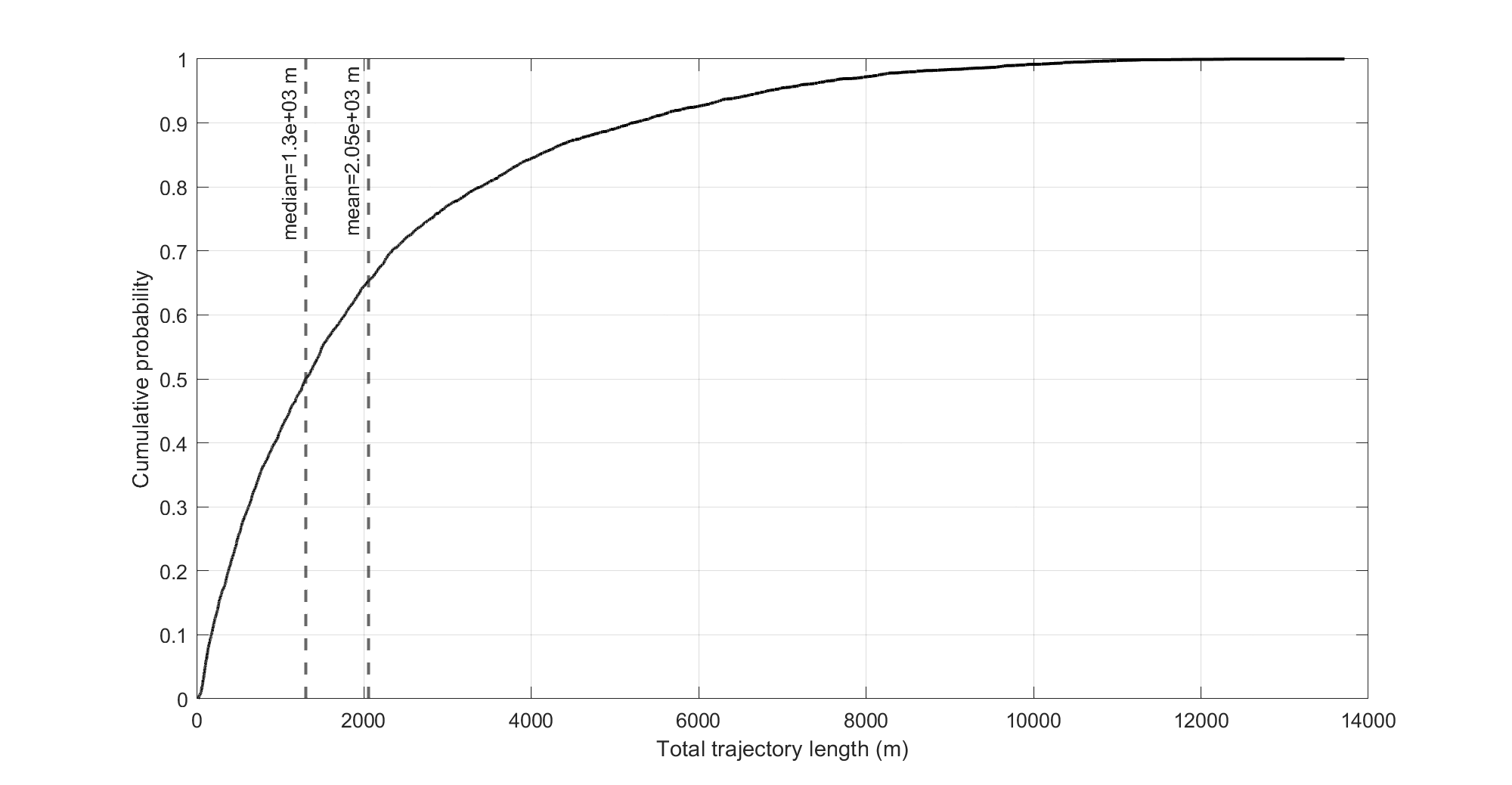}}
\caption{Cumulative distribution function (CDF) of total trajectory length $D_i$. The distribution is strongly right-skewed, with a median $D_i \simeq 1.3\times 10^3$~m and a long tail reaching $\sim 1.4\times 10^4$~m, indicating that a significant amount of starting locations produce kilometre-scale transport.}
  \label{fig:trajlength_cdf}
\end{figure*}

\begin{figure*}

\centerline{\includegraphics[width=0.99\textwidth]{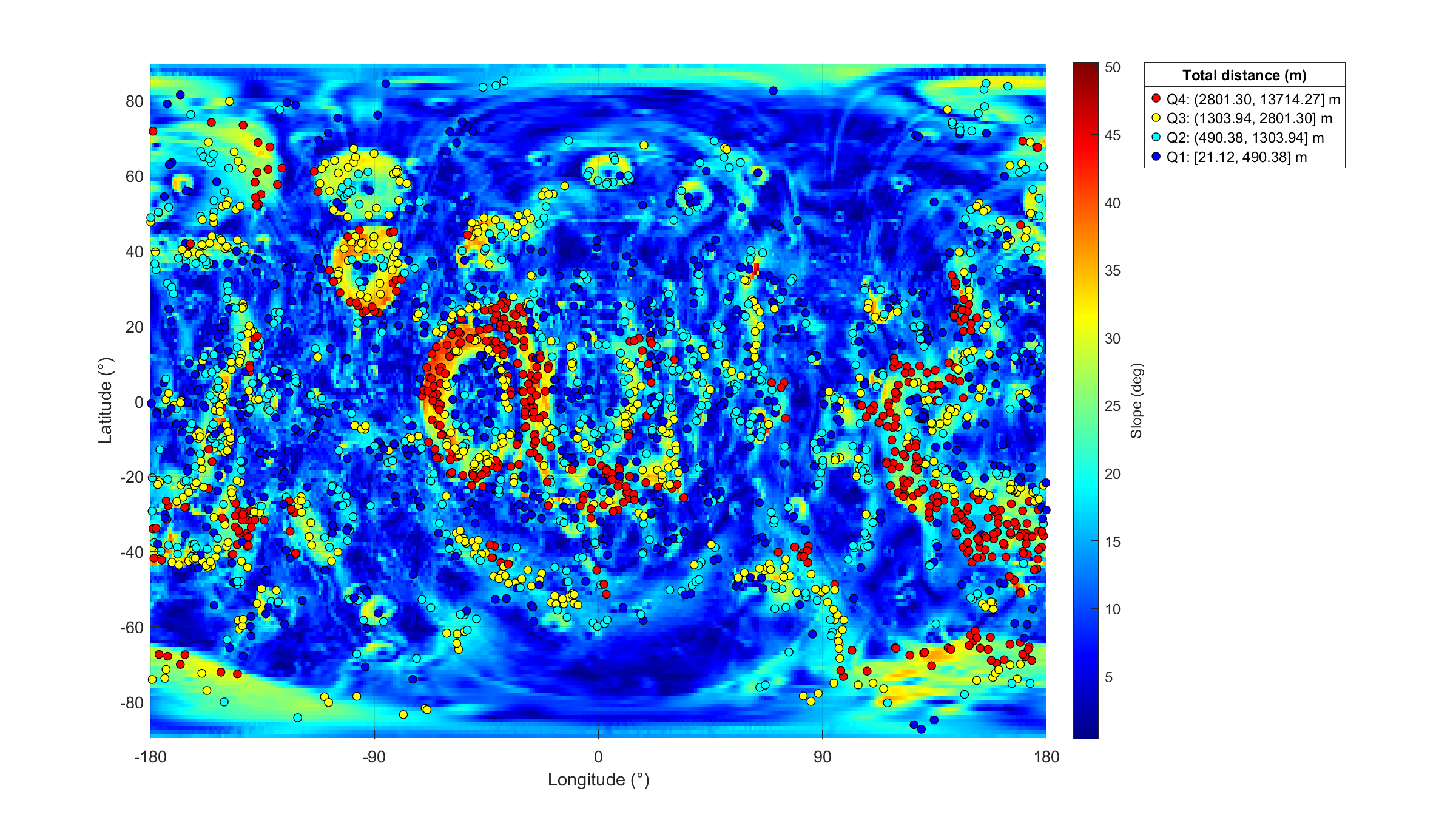}}
\caption{A-posteriori trajectory-length based mobility. Starting points of the surface trajectories are plotted over the dynamical slope map and coloured by quartiles of the total distance travelled along the surface, from blue (shortest paths) to red (longest paths). This map provides a direct measure of how far surface material can be transported from each location, independently of the comparatively minor effect of orbital eccentricity.}
  \label{fig:trajlength_map}
\end{figure*}

\subsection{Comparison of a-priori and a-posteriori maps}\label{Xsec31-C.3}

The a-priori mobility index $M_i$ and the a-posteriori trajectory diagnostics provide complementary views of regolith transport on Phobos. The index $M_i$ is purely local and Eulerian: it depends only on the shape model, the representative acceleration fields (gravity,\break centrifugal and tidal components), and the adopted frictional threshold, and can therefore be evaluated at negligible cost compared with full trajectory integrations. In contrast, the total trajectory length $D_i$ is Lagrangian and measures the effective travelled distance along the surface before stopping, integrating the cumulative influence of the acceleration field, Coriolis deflection, and the local topographic\break constraint.

A direct comparison of the maps shows that regions of high $M_i$ systematically coincide with starting locations producing the longest trajectories (upper quartiles of $D_i$; \linkref[Fig.]{\ref{fig:trajlength_map}}). This spatial agreement confirms that the dynamical-slope-based index is an efficient first-order predictor of far-travelling, long-lived surface motion, and validates its use as a rapid classifier of dynamically active versus accumulation-prone terrains.

The eccentricity--effect index $E_i$ captures a different aspect of the problem: it quantifies how strongly the peri-/apo-center dynamical conditions alter the deposition endpoints for material launched from the same starting location. Using paired peri-/apo-center trajectories, we obtain finite values spanning $E_i \in [3.6\times10^{-4},\,1.91]$, with a strongly skewed distribution dominated by small values (\linkref[Fig.]{\ref{CDF_ecc}}). In particular, the cumulative distribution rises steeply, indicating that most starting locations experience only modest endpoint shifts relative to the typical excursion distance, while a minority of cases populate a long tail toward $E_i\sim\mathcal{O}(1)$.

Spatially, the highest $E_i$ quartiles occur predominantly within regions that are already dynamically active according to $M_i$ and $D_i$, consistent with the fact that endpoint sensitivity can only manifest where non-negligible motion exists. However, the $E_i$ quartile map (\linkref[Fig.]{\ref{fig:ecc_map}}) is noticeably more heterogeneous than the $D_i$ map, suggesting that orbital eccentricity acts as a secondary modulator of endpoint dispersion within otherwise mobile terrains, rather than as a primary control on where transport occurs.

Taken together, these diagnostics indicate that the global pattern of regolith mobility is dominated by the shape-controlled effective gravity field and local topography, as captured by $M_i$ and confirmed by the travelled distances $D_i$. Orbital eccentricity does not redefine the main transport domains or the overall Regolith Migration Pathways, but it can introduce a measurable dispersion of deposition endpoints for a subset of starting locations. Therefore, while eccentricity can be neglected for the global mobility patterns addressed in this study, it should be taken into account in local analyses where source-to-sink connectivity is interpreted at the scale of individual \break trajectories.

\appsection{Effect of allowing trajectory detachment from the surface } \label{appendix:flying_particles_dynamics} \label{surface-constraint}

\def\thetable{D.\arabic{table}}
\setcounter{table}{0}
\def\thefigure{D.\arabic{figure}}
\setcounter{figure}{0}

In this section, we use RAVEL in an alternative integration mode to explore trajectories computed without enforcing the constraint that motion remains strictly bound to the surface. The trajectories are integrated following the same formulation as in the surface-constrained case,
except that the acceleration component normal to the local surface is no longer suppressed.
As a result, trajectories are allowed to temporarily detach from the surface and follow short ballistic excursions before re-impacting Phobos' surface.\vfill\pagebreak

During these ballistic phases, the gravitational acceleration exerted  by Phobos is recomputed at every time step using a brute-force \textit{mascon} approach.
In this method, Phobos' gravity field is approximated by summing the individual gravitational contributions from $10^{6}$ equal-mass points randomly and homogeneously distributed throughout the interior of the body.
Although computationally expensive, this technique provides a flexible representation of the gravitational field without assuming a simplified geometry.

Upon re-contact with the surface, trajectories undergo an inelastic rebound characterized by a normal coefficient of restitution set to 0.1 in velocity,
thereby strongly damping vertical motion and promoting rapid reattachment to the surface.
After impact, trajectories resume surface motion governed by the same frictional criteria as in the constrained simulations.

We simulated \del{10000}\ins{10,000} trajectories starting from initial locations distributed randomly over the surface of Phobos,
using the low-threshold  end-member of $14^{\circ}$, corresponding to the mean dynamical slope of the body.
To remain as general as possible, Phobos' orbital motion around Mars is computed simultaneously.
At $t=0$, Phobos is placed at pericenter, and all trajectories are initiated from rest.
The trajectories are integrated over two full Phobos orbital periods, a duration sufficient for all of them to reach a stable final\break state.

The resulting trajectories are shown as black solid lines in \linkref[\del{Figure}\ins{Fig.}]{\ref{fig:appendix_trajectories_flying_part_14_deg_4panels}}
and can be directly compared with the surface-constrained simulations presented in \linkref[\del{Figure}\ins{Fig.}]{\ref{fig:trajectories_14_deg_4panels}}.
Only minor and highly localized differences are observed between the two cases; these are highlighted by orange circles in the figure.
Overall, allowing temporary detachment and ballistic motion does not significantly alter the large-scale pattern of regolith migration,
suggesting that surface-constrained dynamics capture the dominant behavior of regolith transport on Phobos under the conditions considered here.

 \begin{figure*}

\centerline{\includegraphics[width=0.99\textwidth]{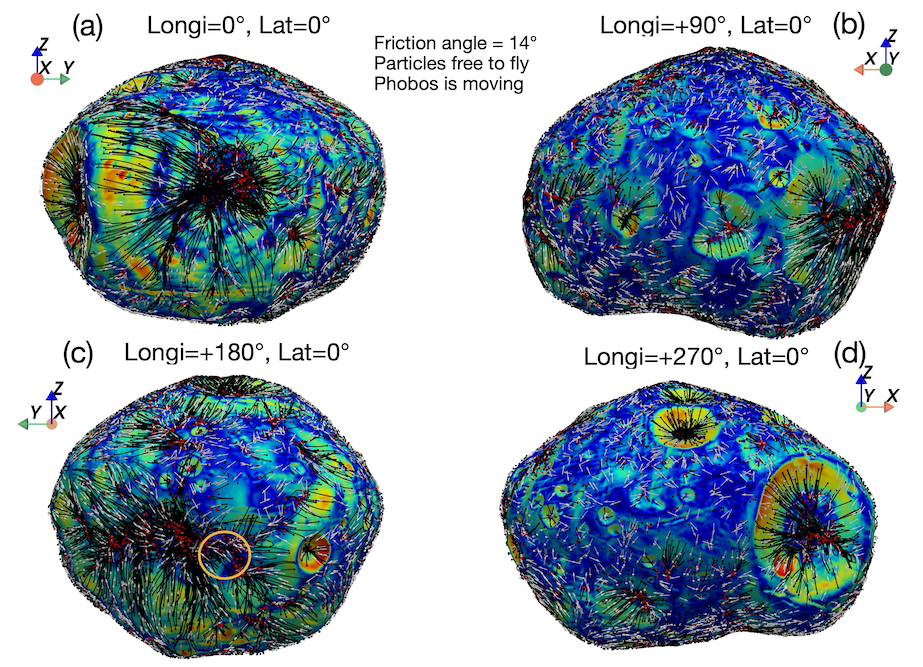}}
\caption{Surface trajectories simulated for $\phi = 14^{\circ}$ when trajectories are allowed to detach from the surface. Phobos is also moving along its current eccentric orbit. Faces at 0, 90, 180, and 270$^{\circ}$ are displayed. Black solid lines are trajectories, red dots are endpoints. Black dots represent starting points. These trajectories are mostly similar to the surface-constrained case. Only regions circled in orange show significant differences between the surface-constrained and unconstrained cases (compare with  \linkref[\del{Figure}\ins{Fig.}]{\ref{fig:trajectories_14_deg_4panels}}).}
    \label{fig:appendix_trajectories_flying_part_14_deg_4panels}
 \end{figure*}

  \begin{figure*}

\centerline{\includegraphics[width=0.95\textwidth]{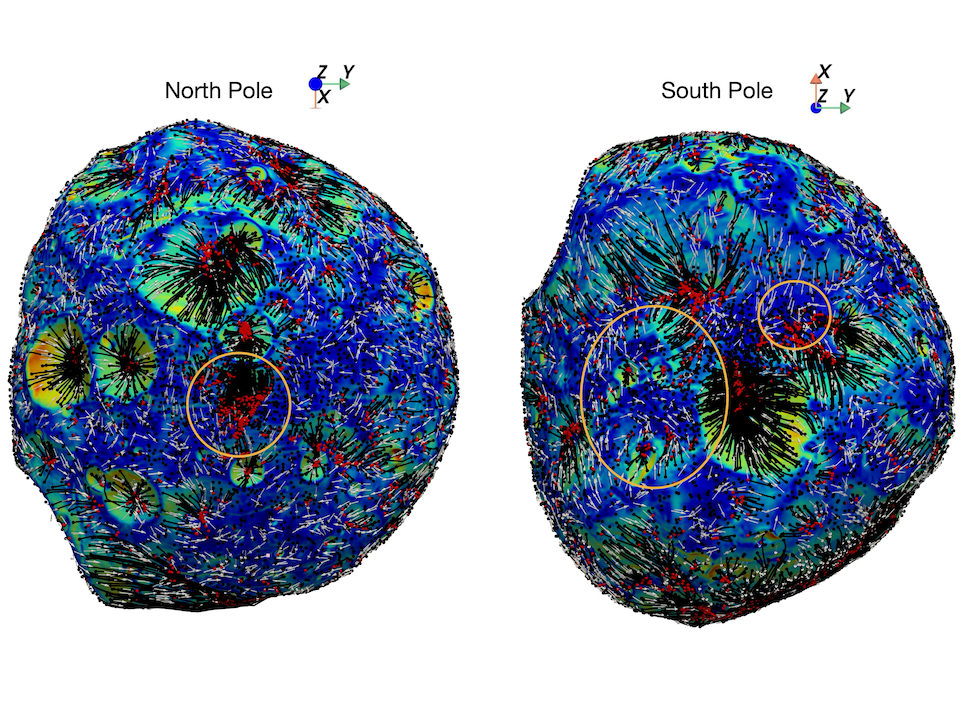}}
\caption{Surface trajectories simulated for $\phi = 14^{\circ}$ when trajectories are allowed to detach from the surface. Phobos is also moving along its current eccentric orbit. Same legend as in \linkref[\del{Figure}\ins{Fig.}]{\ref{fig:appendix_trajectories_flying_part_14_deg_4panels}}. Only regions circled in orange show significant differences between the surface-constrained and unconstrained cases (compare with  \linkref[\del{Figure}\ins{Fig.}]{\ref{fig:trajectories_14_deg_poles}}).}
    \label{fig:appendix_trajectories_flying_part_14_deg_north_south}
 \end{figure*}

\end{appgroup}


\begin{thebibliography}{29}
\expandafter\ifx\csname natexlab\endcsname\relax\def\natexlab#1{#1}\fi
\expandafter\ifx\csname url\endcsname\relax
  \def\url#1{\texttt{#1}}\fi
\expandafter\ifx\csname urlprefix\endcsname\relax\def\urlprefix{URL }\fi
\providecommand{\eprint}[2][]{\url{#2}}
\providecommand{\bibinfo}[2]{#2}
\renewcommand{\newblock}{}
\makeatletter
\ifx\xfnm\relax \def\xfnm[#1]{\unskip,\space#1}\fi
\ifx\xsnm\@undefined\def\xsnm[#1]{#1}\fi
\ifx\plxcitation\@undefined\def\plxcitation#1#2#3#4#5#6{#6}\fi
\ifx\endplxcitation\@undefined\def\endplxcitation{}\fi
\ifx\PrintOrdinal\@undefined\def\PrintOrdinal#1{#1}\fi
\makeatother
\bibitem[{Ballouz {\rm et~al.}(2019)Ballouz, Baresi, Crites, Kawakatsu and
  Fujimoto}]{Ballouz2019}
\plxcitation{}{Ballouz2019}{sbref0001}{}{article}{%
\bibinfo{author}{\xsnm[Ballouz]\xfnm[, R.-L.]},
  \bibinfo{author}{\xsnm[Baresi]\xfnm[, N.]},
  \bibinfo{author}{\xsnm[Crites]\xfnm[, S.~T.]},
  \bibinfo{author}{\xsnm[Kawakatsu]\xfnm[, Y.]},
  \bibinfo{author}{\xsnm[Fujimoto]\xfnm[, M.]}, \bibinfo{year}{2019}.
\newblock \bibinfo{title}{Surface refreshing of martian moon phobos by orbital
  eccentricity-driven grain motion}.
\newblock \bibinfo{journal}{\del{Nature Geoscience}\ins{Nat. Geosci.}} \bibinfo{volume}{12},
  \bibinfo{pages}{229--234}.
\bibinfo{doi}{https://doi.org/10.1038/s41561-019-0323-9}}\endplxcitation
\bibitem[{Barnouin {\rm et~al.}(2019)Barnouin, Daly, Palmer {\rm et~al.}}]{Barnouin2019}
\plxcitation{}{Barnouin2019}{sbref0002}{}{article}{%
\bibinfo{author}{\xsnm[Barnouin]\xfnm[, O.~S.]},
  \bibinfo{author}{\xsnm[Daly]\xfnm[, M.~G.]},
  \bibinfo{author}{\xsnm[Palmer]\xfnm[, E.~E.]}, {\rm et~al.}, \bibinfo{year}{2019}.
\newblock \bibinfo{title}{Shape of (101955) bennu indicative of a rubble pile
  with internal stiffness}.
\newblock \bibinfo{journal}{\del{Nature Geoscience}\ins{Nat. Geosci.}} \bibinfo{volume}{12},
  \bibinfo{pages}{247--252}.
\bibinfo{doi}{https://doi.org/10.1038/s41561-019-0330-x}}\endplxcitation
\bibitem[{Basilevsky {\rm et~al.}(2014)Basilevsky, Lorenz, Shingareva, Head, Ramsley
  and Zubarev}]{Basilevsky_2014}
\plxcitation{}{Basilevsky_2014}{sbref0003}{}{article}{%
\bibinfo{author}{\xsnm[Basilevsky]\xfnm[, A.~T.]},
  \bibinfo{author}{\xsnm[Lorenz]\xfnm[, C.~A.]},
  \bibinfo{author}{\xsnm[Shingareva]\xfnm[, T.~V.]},
  \bibinfo{author}{\xsnm[Head]\xfnm[, J.~W.]},
  \bibinfo{author}{\xsnm[Ramsley]\xfnm[, K.~R.]},
  \bibinfo{author}{\xsnm[Zubarev]\xfnm[, A.~E.]}, \bibinfo{year}{2014}.
\newblock \bibinfo{title}{The surface geology and geomorphology of phobos}.
\newblock \bibinfo{journal}{\del{Planetary and Space Science}\ins{Planet. Space Sci.}} \bibinfo{volume}{102},
  \bibinfo{pages}{95--118}.
\bibinfo{doi}{https://doi.org/10.1016/j.pss.2014.04.013}}\endplxcitation
\bibitem[{C{\'a}rdenas-Barrantes and Ovalle(2025)}]{Cardena-Barrantes2025}
\plxcitation{}{Cardena-Barrantes2025}{sbref0004}{}{article}{%
\bibinfo{author}{\xsnm[C{\'a}rdenas-Barrantes]\xfnm[, M.]},
  \bibinfo{author}{\xsnm[Ovalle]\xfnm[, C.]}, \bibinfo{year}{2025}.
\newblock \bibinfo{title}{Effect of friction on the angle of repose of
  elongated particles}.
\newblock \bibinfo{journal}{\del{Powder Technology}\ins{Powder Technol.}} \bibinfo{volume}{458},
  \bibinfo{pages}{120974}.
\bibinfo{doi}{https://doi.org/10.1016/j.powtec.2025.120974}}\endplxcitation
\bibitem[{Carrier {\rm et~al.}(1991)Carrier, Olhoeft and Mendell}]{Carrier1991}
\plxcitation{}{Carrier1991}{sbref0005}{}{incollection}{%
\bibinfo{author}{\xsnm[Carrier]\xfnm[, W.~D.]},
  \bibinfo{author}{\xsnm[Olhoeft]\xfnm[, G.~R.]},
  \bibinfo{author}{\xsnm[Mendell]\xfnm[, W.]}, \bibinfo{year}{1991}.
\newblock \bibinfo{title}{Physical properties of the lunar surface}.
\newblock In: \bibinfo{editor}{\xsnm[Heiken]\xfnm[, G.~H.]},
  \bibinfo{editor}{\xsnm[Vaniman]\xfnm[, D.~T.]},
  \bibinfo{editor}{\xsnm[French]\xfnm[, B.~M.]} (Eds.),
  \bibinfo{booktitle}{Lunar Sourcebook}. \bibinfo{publisher}{Cambridge
  University Press}, pp. \bibinfo{pages}{475--594}.
}\endplxcitation
\bibitem[{Ernst {\rm et~al.}(2023)Ernst, Daly, Gaskell, Barnouin, Nair, Hyatt,
  Al~Asad and Hoch}]{Ernst_2023}
\plxcitation{}{Ernst_2023}{sbref0006}{}{article}{%
\bibinfo{author}{\xsnm[Ernst]\xfnm[, C.~M.]},
  \bibinfo{author}{\xsnm[Daly]\xfnm[, R.~T.]},
  \bibinfo{author}{\xsnm[Gaskell]\xfnm[, R.~W.]},
  \bibinfo{author}{\xsnm[Barnouin]\xfnm[, O.~S.]},
  \bibinfo{author}{\xsnm[Nair]\xfnm[, H.]},
  \bibinfo{author}{\xsnm[Hyatt]\xfnm[, B.~A.]},
  \bibinfo{author}{\xsnm[Al~Asad]\xfnm[, M.~M.]},
  \bibinfo{author}{\xsnm[Hoch]\xfnm[, K.~K.~W.]}, \bibinfo{year}{2023}.
\newblock \bibinfo{title}{High-resolution shape models of phobos and deimos
  from stereophotoclinometry}.
\newblock \bibinfo{journal}{\del{Earth, Planets and Space}\ins{Earth Planets Space}} \bibinfo{volume}{75},
  \bibinfo{pages}{103}.
\bibinfo{doi}{https://doi.org/10.1186/s40623-023-01814-7}}\endplxcitation
\bibitem[{Fornasier {\rm et~al.}(2024)Fornasier, Wargnier, Hasselmann, Tirsch, Matz,
  Doressoundiram, Gautier and Barucci}]{fornasier2024}
\plxcitation{}{fornasier2024}{sbref0007}{}{article}{%
\bibinfo{author}{\xsnm[Fornasier]\xfnm[, S.]},
  \bibinfo{author}{\xsnm[Wargnier]\xfnm[, A.]},
  \bibinfo{author}{\xsnm[Hasselmann]\xfnm[, P.~H.]},
  \bibinfo{author}{\xsnm[Tirsch]\xfnm[, D.]},
  \bibinfo{author}{\xsnm[Matz]\xfnm[, K.-D.]},
  \bibinfo{author}{\xsnm[Doressoundiram]\xfnm[, A.]},
  \bibinfo{author}{\xsnm[Gautier]\xfnm[, T.]},
  \bibinfo{author}{\xsnm[Barucci]\xfnm[, M.~A.]}, \bibinfo{year}{2024}.
\newblock \bibinfo{title}{Phobos photometric properties from mars express
  {HRSC} observations}.
\newblock \bibinfo{journal}{\del{Astronomy \& Astrophysics}\ins{Astron. Astrophys.}} \bibinfo{volume}{686}
  (\bibinfo{number}{A203}).
\bibinfo{doi}{https://doi.org/10.1051/0004-6361/202449220}}\endplxcitation
\bibitem[{Fraeman {\rm et~al.}(2012)Fraeman, Arvidson, Murchie and
  Rivkin}]{fraeman2012}
\plxcitation{}{fraeman2012}{sbref0008}{}{article}{%
\bibinfo{author}{\xsnm[Fraeman]\xfnm[, A.~A.]},
  \bibinfo{author}{\xsnm[Arvidson]\xfnm[, R.~E.]},
  \bibinfo{author}{\xsnm[Murchie]\xfnm[, S.~L.]},
  \bibinfo{author}{\xsnm[Rivkin]\xfnm[, A.~S.]}, \bibinfo{year}{2012}.
\newblock \bibinfo{title}{Analysis of disk-resolved {OMEGA} and {CRISM}
  spectral observations of phobos and deimos}.
\newblock \bibinfo{journal}{\del{Journal of Geophysical Research: Planets}\ins{J. Geophys. Res. Planets}}
  \bibinfo{volume}{117}, \bibinfo{pages}{E00J15}.
\bibinfo{doi}{https://doi.org/10.1029/2012JE004137}}\endplxcitation
\bibitem[{Fraeman {\rm et~al.}(2014)Fraeman, Murchie, Arvidson {\rm et~al.}}]{fraeman2014}
\plxcitation{}{fraeman2014}{sbref0009}{}{article}{%
\bibinfo{author}{\xsnm[Fraeman]\xfnm[, A.~A.]},
  \bibinfo{author}{\xsnm[Murchie]\xfnm[, S.~L.]},
  \bibinfo{author}{\xsnm[Arvidson]\xfnm[, R.~E.]}, {\rm et~al.},
  \bibinfo{year}{2014}.
\newblock \bibinfo{title}{Spectral absorptions on phobos and deimos in the
  visible/near infrared wavelengths and their compositional constraints}.
\newblock \bibinfo{journal}{\del{Icarus}\ins{ICARUS}} \bibinfo{volume}{229},
  \bibinfo{pages}{196--205}.
\bibinfo{doi}{https://doi.org/10.1016/j.icarus.2013.11.021}}\endplxcitation
\bibitem[{Jacobson and Lainey(2014)}]{Jacobson_Lainey_2014}
\plxcitation{}{Jacobson_Lainey_2014}{sbref0010}{}{article}{%
\bibinfo{author}{\xsnm[Jacobson]\xfnm[, R.~A.]},
  \bibinfo{author}{\xsnm[Lainey]\xfnm[, V.]}, \bibinfo{year}{2014}.
\newblock \bibinfo{title}{Martian satellite orbits and ephemerides}.
\newblock \bibinfo{journal}{\del{Planetary and Space Science}\ins{Planet. Space Sci.}} \bibinfo{volume}{102},
  \bibinfo{pages}{35--44}.
\bibinfo{doi}{https://doi.org/10.1016/j.pss.2013.06.003}}\endplxcitation
\bibitem[{Johnson {\rm et~al.}(2016)Johnson, Campbell and
  Melosh}]{Johnson_2016_low_friction_angle}
\plxcitation{}{Johnson_2016_low_friction_angle}{sbref0011}{}{article}{%
\bibinfo{author}{\xsnm[Johnson]\xfnm[, B.~C.]},
  \bibinfo{author}{\xsnm[Campbell]\xfnm[, C.~S.]},
  \bibinfo{author}{\xsnm[Melosh]\xfnm[, H.~J.]}, \bibinfo{year}{2016}.
\newblock \bibinfo{title}{The reduction of friction in long runout landslides
  as an emergent phenomenon}.
\newblock \bibinfo{journal}{\del{Journal of Geophysical Research: Earth Surface}\ins{J. Geophys. Res. Earth Surf.}}
  \bibinfo{volume}{121} (\bibinfo{number}{5}), \bibinfo{pages}{881--889}.
\bibinfo{doi}{https://doi.org/10.1002/2015JF003751}}\endplxcitation
\bibitem[{Karachevtseva {\rm et~al.}(2014)Karachevtseva, Oberst, Zubarev, Nadezhdina,
  Kokhanov, Garov, Uchaev, Uchaev, Malinnikov and
  Klimkin}]{Karachevtseva_2014_Phobos_Trek}
\plxcitation{}{Karachevtseva_2014_Phobos_Trek}{sbref0012}{}{article}{%
\bibinfo{author}{\xsnm[Karachevtseva]\xfnm[, I.~P.]},
  \bibinfo{author}{\xsnm[Oberst]\xfnm[, J.]},
  \bibinfo{author}{\xsnm[Zubarev]\xfnm[, A.~E.]},
  \bibinfo{author}{\xsnm[Nadezhdina]\xfnm[, I.~E.]},
  \bibinfo{author}{\xsnm[Kokhanov]\xfnm[, A.~A.]},
  \bibinfo{author}{\xsnm[Garov]\xfnm[, A.~S.]},
  \bibinfo{author}{\xsnm[Uchaev]\xfnm[, D.~V.]},
  \bibinfo{author}{\xsnm[Uchaev]\xfnm[, Dm.~V.]},
  \bibinfo{author}{\xsnm[Malinnikov]\xfnm[, V.~A.]},
  \bibinfo{author}{\xsnm[Klimkin]\xfnm[, N.~D.]}, \bibinfo{year}{2014}.
\newblock \bibinfo{title}{The phobos information system}.
\newblock \bibinfo{journal}{\del{Planetary and Space Science}\ins{Planet. Space Sci.}} \bibinfo{volume}{102},
  \bibinfo{pages}{74--85}.
\bibinfo{doi}{https://doi.org/10.1016/j.pss.2013.12.015}}\endplxcitation
\bibitem[{{Le~Maistre} {\rm et~al.}(2019){Le~Maistre}, {Rivoldini} and
  {Rosenblatt}}]{Le_Maistre_2019}
\plxcitation{}{Le_Maistre_2019}{sbref0013}{}{article}{%
\bibinfo{author}{\xsnm[{Le~Maistre}]\xfnm[, S.]},
  \bibinfo{author}{\xsnm[{Rivoldini}]\xfnm[, A.]},
  \bibinfo{author}{\xsnm[{Rosenblatt}]\xfnm[, P.]}, \bibinfo{year}{2019}.
\newblock \bibinfo{title}{{Signature of Phobos' interior structure in its
  gravity field and libration}}.
\newblock \bibinfo{journal}{\del{Icarus}\ins{ICARUS}} \bibinfo{volume}{321},
  \bibinfo{pages}{272--290}.
\bibinfo{doi}{https://doi.org/10.1016/j.icarus.2018.11.022}}\endplxcitation
\bibitem[{{Le~Maistre} {\rm et~al.}(2013){Le~Maistre}, {Rosenblatt}, {Rambaux},
  {Castillo-Rogez}, {Dehant} and {Marty}}]{Le_Maistre_2013}
\plxcitation{}{Le_Maistre_2013}{sbref0014}{}{article}{%
\bibinfo{author}{\xsnm[{Le~Maistre}]\xfnm[, S.]},
  \bibinfo{author}{\xsnm[{Rosenblatt}]\xfnm[, P.]},
  \bibinfo{author}{\xsnm[{Rambaux}]\xfnm[, N.]},
  \bibinfo{author}{\xsnm[{Castillo-Rogez}]\xfnm[, J.~C.]},
  \bibinfo{author}{\xsnm[{Dehant}]\xfnm[, V.]},
  \bibinfo{author}{\xsnm[{Marty}]\xfnm[, J.-C.]}, \bibinfo{year}{2013}.
\newblock \bibinfo{title}{{{Phobos interior from librations determination using
  Doppler and star tracker measurements}}}.
\newblock \bibinfo{journal}{\del{Planetary and Space Science}\ins{Planet. Space Sci.}} \bibinfo{volume}{85},
  \bibinfo{pages}{106--122}.
\bibinfo{doi}{https://doi.org/10.1016/j.pss.2013.06.015}}
\bibitem[{Lei {\rm et~al.}(2025)Lei, Mao and Yu}]{Lei_2025}
\plxcitation{}{Lei_2025}{sbref0015}{}{article}{%
\bibinfo{author}{\xsnm[Lei]\xfnm[, Z.]}, \bibinfo{author}{\xsnm[Mao]\xfnm[,
  W.]}, \bibinfo{author}{\xsnm[Yu]\xfnm[, F.]}, \bibinfo{year}{2025}.
\newblock \bibinfo{title}{{Dynamics of long-runout landslides: \del{A}\ins{a} review}}.
\newblock \bibinfo{journal}{\del{Applied Sciences}\ins{Appl. Sci.}} \bibinfo{volume}{15}
  (\bibinfo{number}{21}).
\bibinfo{doi}{https://doi.org/10.3390/app152111300}}\endplxcitation
\bibitem[{{Lucas} {\rm et~al.}(2014){Lucas}, {Mangeney} and {Ampuero}}]{Lucas_2024}
\plxcitation{}{Lucas_2024}{sbref0016}{}{article}{%
\bibinfo{author}{\xsnm[{Lucas}]\xfnm[, A.]},
  \bibinfo{author}{\xsnm[{Mangeney}]\xfnm[, A.]},
  \bibinfo{author}{\xsnm[{Ampuero}]\xfnm[, J.~P.]}, \bibinfo{year}{2014}.
\newblock \bibinfo{title}{{{Frictional velocity-weakening in landslides on Earth
  and on other planetary bodies}}}.
\newblock \bibinfo{journal}{\del{Nature Communications}\ins{Nat. Commun.}} \bibinfo{volume}{5},
  \bibinfo{pages}{3417}.
\bibinfo{doi}{https://doi.org/10.1038/ncomms4417}}\endplxcitation
\bibitem[{Mitchell {\rm et~al.}(1972)Mitchell, Scott, Houston {\rm et~al.}}]{mitchell1972}
\plxcitation{}{mitchell1972}{sbref0017}{}{inproceedings}{%
\bibinfo{author}{\xsnm[Mitchell]\xfnm[, J.~K.]},
  \bibinfo{author}{\xsnm[Scott]\xfnm[, R.~F.]},
  \bibinfo{author}{\xsnm[Houston]\xfnm[, W.~N.]}, {\rm et~al.}, \bibinfo{year}{1972}.
\newblock \bibinfo{title}{{Mechanical properties of lunar soil: \del{D}\ins{d}ensity,
  porosity, cohesion, and angle of internal friction}}.
\newblock In: \bibinfo{booktitle}{Proceedings of the Lunar Science Conference}.
  \bibinfo{volume}{Vol.~3}, pp. \bibinfo{pages}{3235--3253}.
}\endplxcitation
\bibitem[{Miyamoto {\rm et~al.}(2021)Miyamoto, Niihara, Wada, Ogawa, Senshu, Michel
  {\rm et~al.}}]{Miyamoto_2021}
\plxcitation{}{Miyamoto_2021}{sbref0018}{}{article}{%
\bibinfo{author}{\xsnm[Miyamoto]\xfnm[, H.]},
  \bibinfo{author}{\xsnm[Niihara]\xfnm[, T.]},
  \bibinfo{author}{\xsnm[Wada]\xfnm[, K.]},
  \bibinfo{author}{\xsnm[Ogawa]\xfnm[, K.]},
  \bibinfo{author}{\xsnm[Senshu]\xfnm[, H.]},
  \bibinfo{author}{\xsnm[Michel]\xfnm[, P.]}, {\rm et~al.}, \bibinfo{year}{2021}.
\newblock \bibinfo{title}{Surface environment of phobos and phobos simulant
  {UTPS}}.
\newblock \bibinfo{journal}{\del{Earth, Planets and Space}\ins{Earth Planets Space}} \bibinfo{volume}{73},
  \bibinfo{pages}{214}.
\bibinfo{doi}{https://doi.org/10.1186/s40623-021-01406-3}}\endplxcitation
\bibitem[{Rambaux {\rm et~al.}(2012)Rambaux, Castillo-Rogez, Le~Maistre and
  Rosenblatt}]{Rambaux_2012}
\plxcitation{}{Rambaux_2012}{sbref0019}{}{article}{%
\bibinfo{author}{\xsnm[Rambaux]\xfnm[, N.]},
  \bibinfo{author}{\xsnm[Castillo-Rogez]\xfnm[, J.-C.]},
  \bibinfo{author}{\xsnm[Le~Maistre]\xfnm[, S.]},
  \bibinfo{author}{\xsnm[Rosenblatt]\xfnm[, P.]}, \bibinfo{year}{2012}.
\newblock \bibinfo{title}{Rotational motion of phobos}.
\newblock \bibinfo{journal}{\del{Astronomy and Astrophysics}\ins{Astron. Astrophys.}} \bibinfo{volume}{548},
  \bibinfo{pages}{A14}.
\bibinfo{doi}{https://doi.org/10.1051/0004-6361/201219710}}\endplxcitation
\bibitem[{Ramsley and Head(2019)}]{Ramsley_2019}
\plxcitation{}{Ramsley_2019}{sbref0020}{}{article}{%
\bibinfo{author}{\xsnm[Ramsley]\xfnm[, K.~R.]},
  \bibinfo{author}{\xsnm[Head]\xfnm[, J.~W.]}, \bibinfo{year}{2019}.
\newblock \bibinfo{title}{{Origin of phobos grooves: \del{T}\ins{t}esting the stickney crater
  ejecta model}}.
\newblock \bibinfo{journal}{\del{Planetary and Space Science}\ins{Planet. Space Sci.}} \bibinfo{volume}{165},
  \bibinfo{pages}{137--147}.
\bibinfo{doi}{https://doi.org/10.1016/j.pss.2018.11.004}}\endplxcitation
\bibitem[{Ramsley and Head(2021)}]{ramsley2021}
\plxcitation{}{ramsley2021}{sbref0021}{}{article}{%
\bibinfo{author}{\xsnm[Ramsley]\xfnm[, K.~R.]},
  \bibinfo{author}{\xsnm[Head]\xfnm[, J.~W.]}, \bibinfo{year}{2021}.
\newblock \bibinfo{title}{{The origins and geological histories of deimos and
  phobos: \del{H}\ins{h}ypotheses and open questions}}.
\newblock \bibinfo{journal}{\del{Space Science Reviews}\ins{Space Sci. Rev.}} \bibinfo{volume}{217}
  (\bibinfo{number}{86}).
\bibinfo{doi}{https://doi.org/10.1007/s11214-021-00864-1}}\endplxcitation
\bibitem[{Robin {\rm et~al.}(2024)Robin, Duch{\^e}ne, Murdoch {\rm et~al.}}]{robin2024}
\plxcitation{}{robin2024}{sbref0022}{}{article}{%
\bibinfo{author}{\xsnm[Robin]\xfnm[, C.~Q.]},
  \bibinfo{author}{\xsnm[Duch{\^e}ne]\xfnm[, A.]},
  \bibinfo{author}{\xsnm[Murdoch]\xfnm[, N.]}, {\rm et~al.}, \bibinfo{year}{2024}.
\newblock \bibinfo{title}{Mechanical properties of rubble pile asteroids
  (dimorphos, itokawa, ryugu, and bennu) through surface boulder morphological
  analysis}.
\newblock \bibinfo{journal}{\del{Nature Communications}\ins{Nat. Commun.}} \bibinfo{volume}{15},
  \bibinfo{pages}{6203}.
\bibinfo{doi}{https://doi.org/10.1038/s41467-024-50147-w}}\endplxcitation
\bibitem[{{Salamuni{\'c}car} {\rm et~al.}(2014){Salamuni{\'c}car},
  {Lon{\v{c}}ari{\'c}}, {Pina}, {Bandeira} and {Saraiva}}]{Salamuniccar_2014}
\plxcitation{}{Salamuniccar_2014}{sbref0023}{}{article}{%
\bibinfo{author}{\xsnm[{Salamuni{\'c}car}]\xfnm[, G.]},
  \bibinfo{author}{\xsnm[{Lon{\v{c}}ari{\'c}}]\xfnm[, S.]},
  \bibinfo{author}{\xsnm[{Pina}]\xfnm[, P.]},
  \bibinfo{author}{\xsnm[{Bandeira}]\xfnm[, L.]},
  \bibinfo{author}{\xsnm[{Saraiva}]\xfnm[, J.]}, \bibinfo{year}{2014}.
\newblock \bibinfo{title}{{{Integrated method for crater detection from
  topography and optical images and the new {PH}9224{GT} catalogue of Phobos
  impact craters}}}.
\newblock \bibinfo{journal}{\del{Advances in Space Research}\ins{Adv. Space Res.}} \bibinfo{volume}{53}
  (\bibinfo{number}{12}), \bibinfo{pages}{1798--1809}.
\bibinfo{doi}{https://doi.org/10.1016/j.asr.2013.11.006}}\endplxcitation
\bibitem[{Scheeres {\rm et~al.}(2010)Scheeres, Hartzell, S{\'a}nchez and
  Swift}]{Scheeres_2010}
\plxcitation{}{Scheeres_2010}{sbref0024}{}{article}{%
\bibinfo{author}{\xsnm[Scheeres]\xfnm[, D.~J.]},
  \bibinfo{author}{\xsnm[Hartzell]\xfnm[, C.~M.]},
  \bibinfo{author}{\xsnm[S{\'a}nchez]\xfnm[, P.]},
  \bibinfo{author}{\xsnm[Swift]\xfnm[, M.]}, \bibinfo{year}{2010}.
\newblock \bibinfo{title}{{Scaling forces to asteroid surfaces: \del{T}\ins{t}he role of
  cohesion}}.
\newblock \bibinfo{journal}{\del{Icarus}\ins{ICARUS}} \bibinfo{volume}{210}
  (\bibinfo{number}{2}), \bibinfo{pages}{968--984}.
\bibinfo{doi}{https://doi.org/10.1016/j.icarus.2010.07.009}}\endplxcitation
\bibitem[{Shi {\rm et~al.}(2013)Shi, Willner and Oberst}]{shi2013}
\plxcitation{}{shi2013}{sbref0025}{}{inproceedings}{%
\bibinfo{author}{\xsnm[Shi]\xfnm[, X.]}, \bibinfo{author}{\xsnm[Willner]\xfnm[,
  K.]}, \bibinfo{author}{\xsnm[Oberst]\xfnm[, J.]}, \bibinfo{year}{2013}.
\newblock \bibinfo{title}{Evolution of phobos' orbit, tidal forces, dynamical
  topography, and related surface modification processes}.
\newblock In: \bibinfo{booktitle}{{44th Lunar and Planetary Science Conference}}.
\newblock \bibinfo{note}{{Abstract 1889}}.
}\endplxcitation
\bibitem[{Smith {\rm et~al.}(2018)Smith, Edwards, Mommert, Trilling and
  Glotch}]{Smith2018}
\plxcitation{}{Smith2018}{sbref0026}{}{inproceedings}{%
\bibinfo{author}{\xsnm[Smith]\xfnm[, N.]},
  \bibinfo{author}{\xsnm[Edwards]\xfnm[, C.]},
  \bibinfo{author}{\xsnm[Mommert]\xfnm[, M.]},
  \bibinfo{author}{\xsnm[Trilling]\xfnm[, D.]},
  \bibinfo{author}{\xsnm[Glotch]\xfnm[, T.]}, \bibinfo{year}{2018}.
\newblock \bibinfo{title}{Mapping the thermal inertia of phobos using
  {MGS}-{TES} observations and thermophysical modeling}.
\newblock In: \bibinfo{booktitle}{Lunar and Planetary Science Conference}.
  \bibinfo{volume}{Vol.~49}.
\newblock \bibinfo{note}{{LPI Contribution No. 2083}}.
}\endplxcitation
\bibitem[{Thomas {\rm et~al.}(2011)Thomas, Stelter, Ivanov, Bridges, Herkenhoff and
  McEwen}]{thomas2011}
\plxcitation{}{thomas2011}{sbref0027}{}{article}{%
\bibinfo{author}{\xsnm[Thomas]\xfnm[, N.]},
  \bibinfo{author}{\xsnm[Stelter]\xfnm[, R.]},
  \bibinfo{author}{\xsnm[Ivanov]\xfnm[, A.]},
  \bibinfo{author}{\xsnm[Bridges]\xfnm[, N.~T.]},
  \bibinfo{author}{\xsnm[Herkenhoff]\xfnm[, K.~E.]},
  \bibinfo{author}{\xsnm[McEwen]\xfnm[, A.~S.]}, \bibinfo{year}{2011}.
\newblock \bibinfo{title}{{Spectral heterogeneity on phobos and deimos: Hi{RISE}
  observations and comparisons to mars pathfinder results}}.
\newblock \bibinfo{journal}{\del{Planetary and Space Science}\ins{Planet. Space Sci.}} \bibinfo{volume}{59}
  (\bibinfo{number}{13}), \bibinfo{pages}{1281--1292}.
\bibinfo{doi}{https://doi.org/10.1016/j.pss.2010.04.018}}\endplxcitation
\bibitem[{Veverka(1978)}]{Veverka78}
\plxcitation{}{Veverka78}{sbref0028}{}{article}{%
\bibinfo{author}{\xsnm[Veverka]\xfnm[, J.]}, \bibinfo{year}{1978}.
\newblock \bibinfo{title}{The surfaces of phobos and deimos}.
\newblock \bibinfo{journal}{\del{Vistas in Astronomy}\ins{Vistas Astron.}} \bibinfo{volume}{22}
  (\bibinfo{number}{2}), \bibinfo{pages}{163--192}.
\bibinfo{doi}{https://doi.org/10.1016/0083-6656(78)90014-4}}\endplxcitation
\bibitem[{W{\"a}hlisch {\rm et~al.}(2010)W{\"a}hlisch, Willner, Oberst, Matz,
  Scholten, Roatsch, Hoffmann, Semm and Neukum}]{Wahlisch_2010}
\plxcitation{}{Wahlisch_2010}{sbref0029}{}{article}{%
\bibinfo{author}{\xsnm[W{\"a}hlisch]\xfnm[, M.]},
  \bibinfo{author}{\xsnm[Willner]\xfnm[, K.]},
  \bibinfo{author}{\xsnm[Oberst]\xfnm[, J.]},
  \bibinfo{author}{\xsnm[Matz]\xfnm[, K.-D.]},
  \bibinfo{author}{\xsnm[Scholten]\xfnm[, F.]},
  \bibinfo{author}{\xsnm[Roatsch]\xfnm[, T.]},
  \bibinfo{author}{\xsnm[Hoffmann]\xfnm[, H.]},
  \bibinfo{author}{\xsnm[Semm]\xfnm[, S.]},
  \bibinfo{author}{\xsnm[Neukum]\xfnm[, G.]}, \bibinfo{year}{2010}.
\newblock \bibinfo{title}{A new topographic image atlas of phobos}.
\newblock \bibinfo{journal}{\del{Earth and Planetary Science Letters}\ins{Earth Planet. Sci. Lett.}}
  \bibinfo{volume}{294}, \bibinfo{pages}{547--553}.
\bibinfo{doi}{https://doi.org/10.1016/j.epsl.2009.11.003}}\endplxcitation\vfill\pagebreak

\end{thebibliography}
\end{document}